\documentclass[letterpaper, 12pt]{article}
\usepackage[utf8]{inputenc}
\usepackage[margin=1.0in]{geometry}
\usepackage{amsmath}
\usepackage{amssymb}
\usepackage{listings}
\usepackage{caption}
\usepackage{subcaption}
\usepackage{float}
\usepackage{amsmath}
\usepackage{dcolumn}

\usepackage{amsmath}

\usepackage{graphicx}
\usepackage{epstopdf}
\usepackage[justification=centering]{caption}
\usepackage{subcaption}
\usepackage{multirow}
\usepackage{booktabs}
\usepackage{lineno}
\usepackage{float}
\usepackage{setspace}
\usepackage{overpic}
\usepackage{rotating}
\usepackage{wrapfig,lipsum}
\usepackage{mathtools}
\usepackage{bm}
\usepackage[flushleft]{threeparttable}
\usepackage{hyperref}
\hypersetup{
    colorlinks=true,
    linkcolor=blue,
    filecolor=magenta,      
    urlcolor=cyan,
}

\usepackage[dvipsnames]{xcolor}

\usepackage{tabulary}

\usepackage{algorithm}
\usepackage{algpseudocode}

\usepackage{natbib}
\usepackage{xcolor}

\definecolor{mauve}{RGB}{224,176,255}
\newcolumntype{d}[1]{D{.}{.}{#1}}
\usepackage{authblk}

\title{MultiObjMatch: Matching with Optimal Tradeoffs between Multiple Objectives in R}

\author{
    Shichao Han\footnote{To whom correspondence should be addressed. Work done while at University of California, Berkeley. (schan21@berkeley.edu)} and 
    Samuel D. Pimentel\footnote{University of California, Berkeley (spi@berkeley.edu)}
}

\begin{document}
\maketitle

\begin{abstract}
In an observational study, matching aims to create many small sets of similar treated and control units from initial samples that may differ substantially in order to permit more credible causal inferences.  The problem of constructing matched sets may be formulated as an optimization problem, but it can be challenging to specify a single objective function that adequately captures all the design considerations at work.  One solution, proposed by \citet{pimentel2019optimal} is to explore a family of matched designs that are Pareto optimal for multiple objective functions.  We present an R package, \href{https://github.com/ShichaoHan/MultiObjMatch}{\texttt{MultiObjMatch}}, that implements this multi-objective matching strategy using a network flow algorithm for several common design goals: marginal balance on important covariates, size of the matched sample, and average within-pair multivariate distances. We demonstrate the package's flexibility in exploring user-defined tradeoffs of interest via two case studies, a reanalysis of the canonical National Supported Work dataset and a novel analysis of a clinical dataset to estimate the impact of diabetic kidney disease on hospitalization costs.

\end{abstract}

\section{Introduction}

\subsection{Matching as a partial remedy in non-experimental settings}

In principle, randomized experiments are the ideal method for measuring causal effects, since the treated and control groups are similar in the observed covariates and differ only in receipt of treatment. In practice, however, randomized controlled trials are not always possible. 
In non-randomized observational studies, covariates related to the outcome of interest may be imbalanced, or differently distributed in the treated and control groups. Unless some adjustment is made for these confounding covariates, treatment effect estimates will be biased.


Matching is one method for addressing such confounding bias in observational studies.
Matching typically involves finding, for each treated unit (or each member of some subset of the treated group), one or more control units with similar pre-treatment covariates.  This tends to reduce covariate imbalance and make causal inference more credible \citep{rubin1979using}.  Of course, matching can only address bias arising from observed differences in pre-treatment covariates, while randomized experiments are also protective against bias from unobserved differences.  Matching has been widely adopted in epidemiology, political science, psychology and other disciplines \citep{oakes2006propensity,diamond2013genetic,thuong2020impact,morgan2015counterfactuals}.

Many matching methods focus on identifying exact matches on nominal variables, or on minimizing a pre-specified multivariate distance, such as the Mahalanobis distance or the difference in estimated propensity scores, between treated and control units within the same matched set \citep{stuart2010matching, rosenbaum1985constructing, rubin2000combining, rubin2001using}.  However, there is ambiguity about which distance is most important to match on, with some authors advocating against the use of propensity scores entirely \citep{king2019propensity}, and others arguing for combinations of propensity scores with other distances based on outcome models \citep{antonelli2018doubly}.  Furthermore, while emphasizing small absolute differences within matched sets has many statistical benefits, it provides only partial and indirect guarantees of improved covariate balance in the group of treated units and control units selected, and a newer family of matching methods have shifted the primary focus to selecting groups with high aggregate or marginal balance \citep{rosenbaum2007minimum, zubizarreta2012using,pimentel2015large}. Another question arising frequently in the literature concerns settings where most units can be matched easily but a minority are more difficult to match.  This is particularly important when matching is conducted without replacement, meaning that matched sets must be non-overlapping.  Possible solutions include variable-ratio matching and full matching \citep{ming2000substantial, hansen2004full}, which allow some treated units to match to many controls and others to fewer or even several treated units to the same control, or simply excluding a limited number of difficult-to-match individuals entirely
\citep{rosenbaum2012optimal, iacus2012causal, zubizarreta2014matching}.  Full matching offers the potential benefit of incorporating all observations into the design but results in a more complicated structure that requires differential weighting of observations in analysis, while excluding treated units ensures simple unweighted outcome comparisons can be used for analysis.  


Several software packages in R implement matching methods. Many of these packages focus on minimizing multivariate distances between paired subjects. The package \texttt{optmatch} provides tools for constructing and matching on propensity scores and Mahalanobis distances, exact matching on nominal variables, and calipers forbidding large differences on continuous variables, as well as variable-ratio and full matching \citep{hansen2007optmatch, hansen2004full, hansen2019package}. Another package, \texttt{Matching}, 
also offers matching based on a propensity score, a Mahalanobis distance, or a user-specified distance, as well as a genetic algorithm that searches over many weighted multivariate distances to optimize covariate balance.
The \texttt{cem} package implements coarsened exact matching, in which treated and control observations with identical values for coarsened covariates are grouped into strata, with the degree of coarsening chosen by the user
\citep{cem}. The \texttt{bigmatch} package offers distance-based matching with the ability to choose an optimal caliper, or restriction on the maximum distance allowed in any match, to dramatically speed up the computation of optimal matches. \texttt{quickmatch}  implements a form of full matching optimized to handle especially large datasets. Balance-constrained matching is implemented in \texttt{rcbalance}, which allows users to specify a prioritized list of nominal covariates on which to enforce balance and computes solutions using a fast network flow algorithm \citep{pimentel2016large}, and in the companion package \texttt{rcbsubset}, which offers similar functionality with the 
 added ability to exclude difficult-to-match treated subjects.  Another package, \texttt{designmatch}, offers a more flexible set of balance constraints and computes matches using integer programming methods \citep{zubizarreta2018package, bennett2020building}. 
 Finally, the \texttt{MatchIt} package provides a common interface to several of the packages just listed \citep{matchit}.

\subsection{The problem of tradeoffs and the value of a multiobjective approach}

An effective matched design unites several distinct attributes.  First, a matched design aims to balance all important covariates, meaning that the empirical distribution of each covariate among selected controls is similar to the empirical distribution among selected treated units.  This is the same kind of group similarity that a randomized trial achieves, and ensures that certain forms of bias are absent \citep{pimentel2023fine}.  However, it does not guarantee that subjects within any given matched pair have similar values on important variables. It is also desirable to ensure that variables 
be not only marginally balanced but also as similar as possible within pairs, to maximize efficiency and reduce sensitivity to unmeasured bias.  Finally, it is desirable to ensure that the matched design includes as many matched pairs as possible, both to maximize power and to ensure that the study population is clear; if all treated subjects are retained in the match and these treated subjects represent a random sample from some well-defined population the resulting design is relevant for describing treatment effects in that population.  

Unfortunately, explicit optimization for one of these design goals may limit success on others.  For instance, it may be possible to achieve exact balance on a variable that is initially very different between groups by eliminating many difficult-to-match subjects from both groups, but this may lead to an undesirably small design.  Similarly, it may be possible to form matched pairs only between medical patients in the same hospital with near-identical health conditions, but this configuration of pairs may not achieve good balance on demographic factors.  Or it may be possible to match patients closely on some variables but not on others, so that there are tradeoffs between matching closely on two or more multivariate distances that give weight to different covariates.

The aforementioned matching methods, which generally focus on strictly optimizing one design goal while ignoring others,   
may produce matches with poor quality on \textcolor{black}{these} other dimensions. Although certain R packages, such as \texttt{designmatch} and \texttt{rcbalance}, allow some attention to multiple design goals (e.g. balance and within-pair distances) a strict hierarchy is enforced that leads one of these goals to dominate the algorithm's decisions.  
In principle one could test all the different matching methods and compare their results, but the lack of structure or a unifying framework across the different approaches makes it difficult to organize a systematic search across the space of possible matches.

To unify numerous competing frameworks for matching and provide a more flexible method for exploring tradeoffs among different possible objective functions, an explicitly multiobjective view is helpful. In this mode, one solves not for a single objective function enforcing strict emphasis on close pair distance or marginal balance or some other goal, but for Pareto optimal solutions that strike an efficient tradeoff between them. Two recent matching methods proceed in this vein. \citet{rosenbaum2012optimal} considered the tradeoff between pair distances and sample size.  
\citet{king2017balance} optimized for sample size and marginal balance simultaneously, providing an associated R package \texttt{MatchingFrontier} \citep{king2016matchingfrontier}. This latter approach, however, is limited to the case where control units can be reused in multiple matched sets.

\citet{pimentel2019optimal} propose a much more general framework for producing Pareto optimal matches by representing matching as a network flow optimization problem, subsuming both the problems of \citet{rosenbaum2012optimal} and \citet{king2016matchingfrontier} as special cases and allowing substantial flexibility about which objective functions may be traded off. The resulting multiobjective optimization framework for exploring tradeoffs brings structure and algorithmic thinking to the problem of searching over candidate matches, and allows discovery of highly attractive matches that cannot be generated directly by off-the-shelf methods.  
 
 In what follows we review the network-flow-based multiobjective framework of \citet{pimentel2019optimal} from a practical point of view and provide user-friendly software tools, collected in the new R package \texttt{MultiObjMatch} to implement it. Specifically, \texttt{MultiObjMatch} enables users to flexibly specify emphasis on three design goals that can be simultaneously optimized.
 

\subsection{Two observational studies}

To ground our discussion in concrete data analysis, we consider two observational studies in which matching is a natural design choice.  

First, we consider a well-known dataset assessing the impact of government-supported job training on workers' wages, studied originally by \citet{lalonde1986evaluating} and distributed as part of the \texttt{cobalt} package in R \citep{greifer2016covariate}. 
The data set contains 185 subjects from the labor training program National Supported Work Demonstration (NSW) and 429 subjects for comparison from the Population Survey of Income Dynamics (PSID). 
Since the treatment and control groups come from different studies, serious concerns arise about the baseline comparability of groups, and observed differences in variables such as gender, age, race and educational level 
raise concerns about bias in outcome comparisons 
\citep{lalonde1986evaluating, dehejia1999causal}. We consider in detail the tradeoffs inherent among optimizing a pair-wise distance based on all observed covariates, minimizing the number of treated units left unmatched, and minimizing treatment-control imbalance on a race variable. 

     
Secondly, we consider data from the First Affiliated Hospital to Chongqing Medical University, a 3A hospital in China. There are 11,556 hospitalization records for patients from the department of endocrinology, each either diagnosed with diabetic kidney disease (DKD) or not. Researchers at the hospital are interested in estimating the causal effect of diabetic kidney disease on hospitalization cost. However, gender, age, duration of hospitalization, and other complications also affect the hospitalization cost, as identified by previous researchers \citep{cai2018trends, cao2015factors, farshchi2014cost}. Matching DKD patients to those without DKD provides an opportunity to remove covariate imbalance and extract a more compelling effect estimate. In this dataset, the presence of certain treated patients with few comparable controls is highlighted by the tradeoffs approach, which also helps determine how many treated patients must be dropped to achieve adequate balance in key covariates such as age.

\subsection{Outline} 
 
 Section 2 provides an overview of the mathematical representation of matching as a multi-objective optimization problem, including a detailed discussion of relevant objective functions. 
Sections 3 and 4 demonstrate the use of the main functions of the \texttt{MultiObjMatch} package and illustrate the empirical value of examining trade-offs among multiple design goals using the two data examples. Section 5 reviews key takeaways and discusses additional practical issues including hyperparameter selection and computation.

\section{Multi-objective matching: technical background}

  
Before introducing software tools in the context of our two case studies, we review important technical background concepts. The ideas below are treated in much more granular detail in \citet{pimentel2019optimal}.

\subsection{Optimization setup}
\label{subsec:opt}
We focus on the case of pair matching without replacement for ease of exposition.
Suppose there are $n$ total observations $O_i = (Z_i, X_i)$, where $X_{i}$ is a row vector for covariates observed and 
$Z_i$ is a binary treatment variable taking the value of 1 if unit $i$ is assigned treatment, and 0 if unit $i$ is in the control group. We define a match $\mathcal{M}$ as a set of non-overlapping index pairs $(i_{11}, i_{12}), \ldots, (i_{k1}, i_{k2})$  where $i_{j\ell} \in \{1, \ldots, n\}$ for all $j, \ell$.
 The cardinality $|\mathcal{M}|$ of match $\mathcal{M}$ is the number of matched pairs $k$ it contains. Define the set of feasible matches $\mathcal{S} = \{\mathcal{M}: Z_i \neq Z_j \forall (i,j) \in \mathcal{M} \}$; this is the set of all possible matches in which every pair contains exactly one treated unit and one control.   
Let $f_1:\mathcal{S} \longrightarrow \mathbb{R}$, $f_2:\mathcal{S} \longrightarrow \mathbb{R}$ and $f_3:\mathcal{S} \longrightarrow \mathbb{R}$ be three objective functions to be minimized. 

A match $\mathcal{M} \in \mathcal{S}$ is Pareto optimal if no other feasible solution $\mathcal{M}'$ performs equally well on all objective functions, $f_1, f_2, f_3$ and strictly better on at least one.
Solving the following problem is guaranteed to produce a Pareto optimal solution: 


\begin{alignat}{2}
\label{eqn:penalized}
&\!\min_{\mathcal{M}}        &\qquad& \rho_1 f_1(\mathcal{M}) + \rho_2 f_2(\mathcal{M}) + f_3(\mathcal{M})\\
&\text{subject to} &      & \mathcal{M} \in \mathcal{S}\nonumber
\end{alignat}

Here, $\rho_1 > 0$ and $\rho_2 > 0$ are parameters that represent different emphases put on the three objectives. Users specify the value of $\rho$'s based on a preferred emphasis on three design goals. The greater the weight, the more focus is on optimizing for the specific design goal.  


Although the problem allows for three objective functions to be considered simultaneously, it is generally much easier to explore and understand a tradeoff between two objective functions at a time. A good general strategy is to begin by holding one of the tuning parameters either very large or very small and varying the other to examine a specific two-dimensional tradeoff, then moving on to another two-dimensional tradeoff as seems productive.

As discussed in \citet{pimentel2019optimal}, solutions may be obtained whenever the objective functions can be formulated as linear costs in a network flow optimization problem.  However, \texttt{MultiObjMatch} focuses on three specific types of objective function that appear frequently in practice: (1) minimizing the number of treated units left unmatched, (2) minimizing the total variation imbalance on the marginal distribution of key categorical variables, and (3) minimizing sums of one or more within-pair distances. We now discuss each in detail. 

\subsection{Within-pair \textcolor{black}{distances}}
   
One design objective is to ensure units in the same matched pair have similar covariate values.  To quantify this objective, we pre-specify a multivariate distance $d(X_i,X_j) = D_{ij}$   between covariate vectors for any treated unit and any control unit in the study and seek to choose pairs such that distances between paired units are as close to zero as possible. 

\textcolor{black}{Numerous distances have been proposed for matching, each with its own strengths and weaknesses.}
\textcolor{black}{One} common choice, the Mahalanobis distance, 
\textcolor{black}{is} defined as follows:
$$D_{ij} = (X_i - X_j)^{'} \Sigma^{-1} (X_i - X_j)$$  
Here 
$\Sigma$ is the sample covariance matrix of the covariates. 
 Mahalanobis distances are attractive for matching because they adjust well to datasets with different magnitudes of variation and collinearity across covariates  \citep{ho2007matching}, especially compared to the Euclidean distance, which uses an identity matrix in place $\Sigma$. \textcolor{black}{A  robust variant of Mahalanobis distance relies on a robust covariance matrix estimation by replacing each covariate column by its ranks and (when ties are present) altering the diagonal of the covariate covariance matrix to contain variances of ranks without ties \citep[\S 9.3]{rosenbaum2020design}.  This avoids  issues with outliers and with giving undue influence to rare binary variables that can be problematic for the standard Mahalanobis distance. In light of these advantages, our package uses the robust Mahalanobis distance by default in measuring pair-wise distance. }

Distances are also commonly constructed using estimated propensity scores. The population propensity score for a unit $i$ is $e(X_i) = P(Z_i = 1 | X_i = x_i)$, which may be estimated using logistic regression or another method to produce fitted values $\hat{e}(X_i)$. Matching can then be conducted using the distances  $D_{ij} = |\hat{e}(X_i) - \hat{e}(X_j)|$ \citep{austin2008critical, caliendo2008some}. Propensity scores are attractive because they capture all the important signals related to the treatment variable in potentially high-dimensional covariates when correctly specified \citep{rosenbaum1983central}. 

\textcolor{black}{A final distance that has received substantial attention is the prognostic score, which is typically estimated by fitting an outcome model to covariates in a pilot sample consisting entirely of controls \citep{hansen2008prognostic}.  Matching on fitted values from this model removes bias by grouping units with similar expected potential outcomes rather than similar expected treatments, also conferring robustness to model choice and unobserved confounding \citep{rosenbaum2005heterogeneity, king2019propensity}.  Some authors have advocated for matching on a multivariate distance that combines the propensity score and the prognostic score \citep{antonelli2018doubly}.
}
For discussion of other matching distances, see 
 \citet{yu2019directional}.

To summarize the success of this goal across all matched pairs, we let $f_{dist1}$ denote the sum of within-pair distances for the match $\mathcal{M}$ \textcolor{black}{as} $f_{dist1}(\mathcal{M}) = \sum_{(T_k, C_k)\in \mathcal{M}} D_{ij}$.

\subsection{Number of Treated Units Excluded}
The whole idea of pair matching is to improve the quality of a comparison by excluding units from the study.  Whenever the treated group and the control group differ substantially in size in the original sample, however, it is common to exclude only units from the larger group, retaining and matching each of the units in the smaller.  Without loss of generality, we will assume that the treated group is smaller, which is commonly the case in practice.  Nevertheless, there are settings where it may be advantageous to exclude some treated units from the match \citep{rosenbaum2012optimal, resa2016evaluation}. This may be the case either if the groups are initially similar in size, or if some members of the treated group are too dissimilar to anyone in the control sample to be matched satisfactorily. On the other hand, dropping too many units may hurt a study's precision \citep{rubin1996matching, king2016matchingfrontier}.  \textcolor{black}{In addition, dropping units may limit a study's generalizability or representativeness when the original treated sample represents a larger population of interest, since the matched treated sample may describe a  different, ambiguous population \citep{traskin2011defining, bennett2020building}}.  As such, we define the objective function $f_{exclude}({\mathcal{M}})= \sum^n_{i=1}Z_i - |\mathcal{M}|$ as the number of treated units  excluded. 

\subsection{Marginal Balance}  
\label{subsec:margbal}

Minimizing multivariate distances between covariate vectors of paired units can lead to similar overall distributions of covariates in matched samples, but this is not guaranteed.  In particular, many small differences in the same direction across different matched pairs may aggregate into a systematic difference in marginal distributions, or multivariate distances may allow for larger differences on a single variable when the others are close, leading to severe imbalance on that specific variable \citep{rosenbaum2002overt, pimentel2019optimal}.  As such, it is useful to be able to optimize directly for the overall similarity of empirical distributions for a covariate between treated and control groups. 

\textcolor{black}{The most common metric for measuring marginal imbalance is the standardized mean difference.  While minimizing standardized mean differences for all measured covariates would be a logical design goal, this objective function is not well-suited to network flow optimization.  As such,}
we focus on optimizing for \textcolor{black}{ a different measure of marginal imbalance}, the total variation distance between empirical distributions \textcolor{black}{of a single discrete covariate across} the two groups \citep{pimentel2015large}. Formally, let $\mathcal{L}$ be the set of all possible values of a discrete covariate. For any  $\ell 
\in \mathcal{L}$ let $n^t_\ell(\mathcal{M})$ represent the proportion of all treated units contained in match $|\mathcal{M}|$ with that value, and let $n^c_\ell(\mathcal{M})$ be the corresponding number of matched control units.  Then define 
\[
f_{dist2} = \sum_{\ell \in \mathcal{L}}\left|n^t_\ell(\mathcal{M}) - n^c_\ell(\mathcal{M}) \right|.
\]
\textcolor{black}{Technically, this quantity is a rescaling of the total variation distance, which lies on [0,1] and can be obtained by dividing $f_{dist2}$ by twice the number of matched pairs $|\mathcal{M}|$}. Typically this imbalance is constrained for a particular variable of special importance 
 or an interaction of several variables.  For more discussion, see \citet{zubizarreta2012using} and \citet{pimentel2015large}.  \textcolor{black}{We note that minimizing the total variation distance on a single binary covariate is equivalent to minimize the standardized mean difference; otherwise, the two measures of imbalance do not coincide, although they are related and one may easily compute both measures for any given match.  For an alternate optimization approach that can directly optimize multiple standardized mean differences at the cost of higher worst-case computation time, see \citet{bennett2020building}.}

 

\subsection{Additional constraints on matching solutions}
Problem (\ref{eqn:penalized}) considers all matched designs in $\mathcal{S}$, i.e. any choice of non-overlapping pairs that contain exactly one treated unit and one control each.  However, this large set contains many very poor solutions and there are computational and statistical advantages to introducing additional constraints on the optimization space.  One useful tool 
is the caliper \citep{rubin2000combining}. A propensity score caliper rules out any match $\mathcal{M}$ that pairs any two units $i$ and $j$ with $|e_i - e_j| > c$, where $c$ is the user-defined caliper size and $e_i$ and $e_j$ are estimated propensity scores for unit $i$ and $j$.   Rosenbaum and Rubin(1985) suggest a caliper size of 0.25 standard deviations of the linear propensity score \citep{rosenbaum1985constructing}.  Calipers may be imposed on variables besides the estimated propensity score too; when they are imposed on discrete variables with $c=0$, they enforce exact agreement on those variables and are known as exact matching constraints. While these constraints are not the central focus of the tools provided by \texttt{MultiObjMatch}, certain functions can accommodate them, and they are useful in considering how to ensure efficient computation.



\section{Guide to MultiObjMatch: job training example}
\label{sec:lalonde}

\textcolor{black}{We now demonstrate a matching workflow using \texttt{MultiObjMatch} to assess the impact of participation in the NSW program on 1978 wages in the job training data.  As explained below, the particular challenges in this dataset suggest exploring tradeoffs involving marginal balance on participant race.  As such in this section we focus on the \texttt{dist\_bal\_match} function, which permits tradeoffs involving marginal balance. Another important workhorse function \texttt{two\_dist\_match}, which explores tradeoffs between multiple within-pair multivariate distance metrics, is discussed in Appendix \ref{subsec:twodist} using a simulated dataset.}

\subsection{Data pre-processing \textcolor{black}{and initial match}}
\label{subsec:preprocess}

\begin{lstlisting}[language=R]
library(MultiObjMatch)
library(MatchIt)
library(cobalt)

data(lalonde, package = "cobalt")
covariates <- c("age", "educ", "married", "nodegree", 
                "race", "re74", "re75")
treat_val <- "treat"
response_val <- "re78"

initial_match <- matchit(
  treat ~ age + educ + race + married +
    nodegree + re74 + re75,
  data = lalonde,
  method = "optimal", distance = "robust_mahalanobis"
)
\end{lstlisting}

\textcolor{black}{The data frame we load from the \texttt{cobalt} package contains a binary treatment variable \texttt{treat} indicating whether each subject participated in the job training program, an outcome variable \texttt{re78} reporting 1978 wages, and seven covariates: age, years of schooling (\texttt{educ}), race (one of Black, White, and Hispanic), marital status, whether the subject failed to graduate from high school (\texttt{nodegree}), and wages in 1974 (\texttt{re74}) and 1975 (\texttt{re75}).  We note that the dataset does not contain missing values;}
  missing values must be removed from \textcolor{black}{both the treatment indicator and the covariates} prior to running \texttt{MultiObjMatch} commands or errors will be triggered.  For guidance on handling missing values in matching, we refer readers to Chapter 9 of  \citet{rosenbaum2010design}. 
\textcolor{black}{Finally, we use the \texttt{MatchIt} package to conduct optimal matching (via functionality loaded from the \texttt{optmatch} package) on a robust Mahalanobis distance based on all covariates. 
}

\subsection{Tradeoffs between distance, balance, and exclusion} 
\label{subsec:distbal}
Consider \textcolor{black}{the covariate balance attained by the optimal propensity score match} 
(Table \ref{tab:bad_example}).
 Although matching \textcolor{black}{reduces} the \textcolor{black}{initial} imbalance, the marginal distribution of race in both groups remains substantially different.  Given that we might expect correlations between race and income in 1978 (our outcome variable), it is desirable to reduce this imbalance, even at the cost of slightly worse matches on the propensity score.  However, it may not be possible to balance race while retaining all treated units; there are 156 Black subjects in the treated group, but only 87 Black subjects in the entire control pool. Thus it may also be necessary to exclude some of the Black treated subjects in order to bring the race variable into balance. 
The function \texttt{dist\_bal\_match} allows exploration of the tradeoffs among these possibilities.
\begin{table}[h]
\begin{center}

\begin{tabular}{l |r r |r |rr}
\hline
           & \multicolumn{2}{c|}{\textcolor{black}{SMD}}   & \multicolumn{1}{c|}{treated group mean} & \multicolumn{2}{c}{control group mean} \\
           & \multicolumn{1}{c}{before} & \multicolumn{1}{c|}{after}  & \multicolumn{1}{c|}{} & \multicolumn{1}{c}{before} & \multicolumn{1}{c}{after} \\ \hline
age        & -0.309           & -0.152           & 25.82                 & 28.03                    & 226.90          \\
educ       & 0.055            & 0.003         & 10.35                  & 10.24                   & 10.34       \\
race\_black  & 1.762            & 1.056           & 0.84                   & 0.20                   & 0.46          \\
race\_hispan & -0.350            & -0.091           & 0.06                   & 0.14                   & 0.08         \\
race\_white  & -1.882           & -1.222          & 1.00                   & 0.66                   & 0.46         \\
married    & -0.826           & -0.317          & 0.19                   & 0.51                    & 0.31         \\
nodegree   & 0.245            & 0.095          & 0.71                   & 0.60                    & 0.65         \\
re74       & -0.721           & -0.132         & 2095.57                & 5619.24                 & 2742.66      \\
re75       & -0.290            & -0.052          & 1532.06               & 2466.48                 & 1732.96       \\ \hline
\end{tabular}
\end{center}
\caption{\textcolor{black}{Covariate balance before and after optimal robust Mahalanobis distance matching for the job training data, as measured by the standardized mean difference (SMD).} }
\label{tab:bad_example}

\end{table}

\textcolor{black}{As arguments, \texttt{dist\_bal\_match} requires a data frame (\texttt{data}) containing the treatment variable and the covariates, the name of the binary treatment variable (\texttt{treat\_col}) and the name of the discrete covariate on which to enforce marginal balance (\texttt{bal\_val}). Vectors of names for covariates to be used in a robust multivariate Mahalanobis distance must also be supplied (\texttt{dist\_col}); alternatively, a distance matrix can be precomputed and supplied under \texttt{dist\_matrix}.  It is important to remember} 
that the outcome variable should never be used in the pairwise distance or as a target for marginal balance, since this may remove part of the treatment effect we hope to estimate \citep{rosenbaum1984consequences}. 

In addition, \texttt{dist\_bal\_match} requires arguments determining how to explore the tradeoffs. As discussed in Section \ref{subsec:opt}, the key parameters for generating a variety of Pareto optimal solutions are the penalties $\rho_1$ and $\rho_2$ for the number of units excluded and for theimbalance in the marginal balance variable; these are specified as numeric vectors to arguments \texttt{exclude\_penalty} and \texttt{balance\_penalty} respectively. Matches will be computed for all unique combinations of the values in the two vectors.





The code below shows our use of \texttt{dist\_bal\_match} for the job training \textcolor{black}{data for a set of penalty values chosen to illustrate an interesting range of different matches. It is rarely obvious how to choose interesting penalty values initially, and the package provides automated tools to help explore and identify relevant values.  We discuss this process for the current example (as well as the argument \texttt{max\_iter}) below in Section \ref{subsubsec:penalties}.}




\begin{lstlisting}[language=R]
bal_val <- "race"
rho_exclusion <- 94.27
rho_balance <- c(85, 86, 86.5, 87, 87.5, 88, 88.5, 89, 90,
                 91)
match_result <- dist_bal_match(
  lalonde, 
  treat_col = treat_val,
  marg_bal_col = bal_val, 
  exclusion_penalty = rho_exclusion,
  balance_penalty = rho_balance, 
  max_iter = 0
)
\end{lstlisting}



\subsubsection{Displaying and interpreting \texttt{dist\_bal\_match} output}
\label{subsec:display}
\begin{figure}[h!]
     \centering
     \begin{subfigure}[b]{0.45\textwidth}
         \centering
         \includegraphics[width=\textwidth, scale = 1.7]{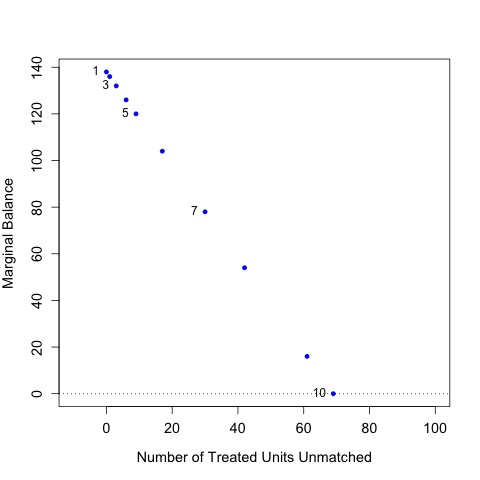}
         \label{fig:tv_unmatch1}
         \caption{}
     \end{subfigure}
     \hfill
     \begin{subfigure}[b]{0.45\textwidth}
         \centering
         \includegraphics[width=\textwidth, scale = 1.7]{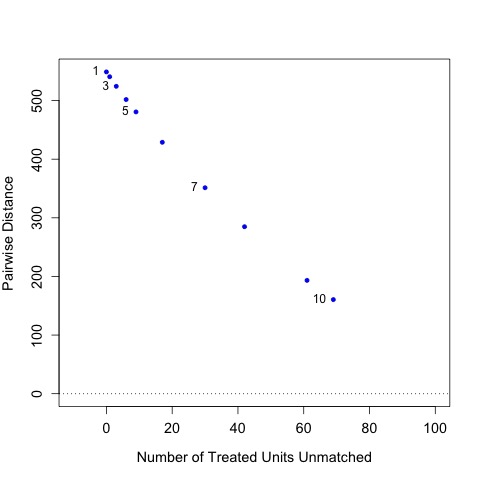}
         \label{fig:pairdist_unmatch1}
         \caption{}
     \end{subfigure}
     \hfill
     \begin{subfigure}[b]{0.45\textwidth}
         \centering
         \includegraphics[width=\textwidth, scale = 1.7]{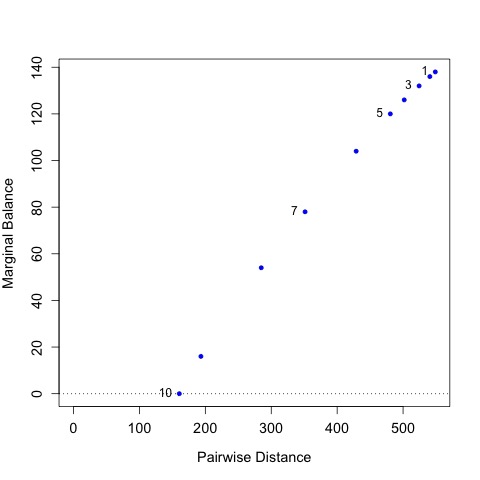}
         \label{fig:pairdist_tv1}
         \caption{}
     \end{subfigure}
        \caption{Three graphs that visualize the trade-offs among three objectives for the \textbf{distBalMatch} example}
        \label{fig:eval_graphs_lalonde1}
\end{figure}

\texttt{dist\_bal\_match} generates a set of matches with varying emphasis on three main objectives, and the next challenge is to quickly understand and visualize the important differences between the resulting matches.  
Tables \ref{table:lalonde-numerical}-\ref{table:percentage} and Figure \ref{fig:eval_graphs_lalonde1}, 
all created by package commands, provide a valuable look at the range of matches just produced for the job training example.

Table \ref{table:lalonde-numerical}, produced by \textcolor{black}{calling  \texttt{summary} on the object produced by \texttt{dist\_bal\_match}}, gives a birds-eye view of the tradeoff in terms of the objective functions and penalty parameters actually used by the underlying network flow algorithm.  Reading the table from top to bottom shows matches with increasingly larger sample sizes, and increasingly worse pair-wise distance and balance (along with the penalty parameters used to produce those matches).  Note that the table contains only a selection of distinct matches computed chosen to illustrate the general shape of the tradeoffs involved, although the R command \textcolor{black}{shows} all distinct matches \textcolor{black}{by default}. \textcolor{black}{We have included raw output from all three calls to \texttt{summary} shown below in Section \ref{sec:console_output} of the Appendix.}

The primary difficulty of Table \ref{table:lalonde-numerical} is that the raw penalty and objective values are not always easily interpretable.  As such we \textcolor{black}{specify \texttt{type = "balance"} when calling \texttt{summary} \ to} generate familiar balance metrics (Table \ref{table:balance_lalonde1}) and \textcolor{black}{use \texttt{type = "exclusion"} to generate} sample size descriptions (Table \ref{table:percentage}). Comparing the standardized differences in Tables \ref{tab:bad_example} and \ref{table:balance_lalonde1}, we see first that all the matches considered improve on the basic \textcolor{black}{optimal match of Section \ref{subsec:preprocess}}.  This suggests that introducing all our design goals into the objective function of our matching problem already tends to improve performance, regardless of the size of the weights associated with each goal.  The other main takeaway from Tables \ref{table:balance_lalonde1}-\ref{table:percentage} is the dramatic improvement in balance realized by excluding some treated units. \textcolor{black}{ In match 1, which excludes no treated units, the standardized differences for Black and White remain larger than 0.35 in magnitude; when we exclude 16.2\% of the treated sample as in match 7, the standardized differences are reduced to 0.252 (Black) and -0.245 (White).  If we are willing to exclude 37.3\% of the sample we can reduce the standardized differences all the way to zero (match 10).
}

\textcolor{black}{A final valuable question to consider is the degree to which the matched treated group differs from the original treated group when treated units are dropped. The \texttt{check\_representative} function computes balance measures comparing these two groups (Table \ref{table:representative_lalonde}).  In match 1, which drops no treated units, all the SMDs are zero, since the matched treated group is identical to the original treated group; the other matches all drop treated units and the groups differ to some degree. Most of the differences are not large, although in match 10, which drops about 3/8 of the treated group, the matched treated group exhibits a markedly higher age and lower pre-treatment income than the overall treated group.}

\begin{lstlisting}
summary(match_result)
summary(match_result, type = "balance")
summary(match_result, type = "exclusion")
check_representative(match_result)
\end{lstlisting} 



\begin{table}[h!]
\centering
\begin{tabular}{l d{1} d{1.5} d{1.1} d{1.5} d{1} d{1}}
\hline
Matching \tnote{a} & \multicolumn{1}{c}{Pair-wise Distance} & \multicolumn{1}{c}{Exclusion} & \multicolumn{1}{c}{Marginal TV} & \multicolumn{1}{c}{Pair-wise} & \multicolumn{1}{c}{Exclusion} & \multicolumn{1}{c}{Marginal TV\tnote{b}}  \\ 
Index & \multicolumn{1}{c}{Penalty} & \multicolumn{1}{c}{Penalty} & \multicolumn{1}{c}{Penalty} & \multicolumn{1}{c}{Distance} & \multicolumn{1}{c}{Cost} & \multicolumn{1}{c}{Distance} \\ \hline
1            & 1       & 94.27      & 85          & 548.870 & 0        & 138       \\
2            & 1       & 94.27      & 86          & 540.612 & 1        & 136       \\
3            & 1       & 94.27      & 86.5        & 524.310 & 3        & 132       \\
4            & 1       & 94.27      & 87          & 501.749 & 6        & 126       \\
5            & 1       & 94.27      & 87.5        & 480.646 & 9        & 120       \\
6            & 1       & 94.27      & 88          & 428.832 & 17       & 104       \\
7            & 1       & 94.27      & 88.5        & 351.303 & 30       & 78        \\
8            & 1       & 94.27      & 89          & 284.819 & 42       & 54        \\
9            & 1       & 94.27      & 90          & 193.292 & 61       & 16        \\
10           & 1       & 94.27      & 91          & 160.513 & 69     & 0       \\ \hline
\end{tabular}


\caption{
 Tabular summary of tradeoffs between balance on race, sample size, and pairwise distance in job training example, via \textcolor{black}{\texttt{summary}}. The numbers are not contiguous, since not all distinct values of the penalty parameter produce distinct matches. ``Exclusion Cost" refers to the number of treated units left unmatched, and columns labeled ``Marginal TV" relate to treatment-control imbalance on race (as measured by the total variation distance).}
\label{table:lalonde-numerical}
\end{table}

\begin{table}[h!]
\centering
\begin{tabular}{l r d{1.3} r d{1.3} d{1.3} d{1.3} d{1.3} d{1.3}}
\hline
             & Before Matching & \multicolumn{1}{c}{Type}    & 10     & 7      & 5      & 3      & 1     \tnote{a} \\
             \hline
age          & -0.242 & \multicolumn{1}{c}{Contin.} & -0.16  & -0.108 & -0.134 & -0.13  & -0.104 \\
married      & -0.324 & \multicolumn{1}{c}{Binary}  & -0.034 & -0.071 & -0.114 & -0.121 & -0.124 \\
educ         & 0.045  & \multicolumn{1}{c}{Contin.} & 0.028  & -0.003 & 0.015  & 0.014  & -0.014 \\
nodegree     & 0.111  & \multicolumn{1}{c}{Binary}  & 0.043  & 0.032  & 0.028  & 0.038  & 0.049  \\
race\_black  & 0.640   & \multicolumn{1}{c}{Binary}  & 0.000      & 0.252  & 0.341  & 0.363  & 0.373  \\
race\_hispan & -0.083 & \multicolumn{1}{c}{Binary}  & 0.000      & -0.006 & -0.006 & -0.016 & -0.022 \\
race\_white  & -0.558 & \multicolumn{1}{c}{Binary}  & 0.000      & -0.245 & -0.335 & -0.346 & -0.351 \\
re74         & -0.596 & \multicolumn{1}{c}{Contin.} & -0.003 & -0.098 & -0.126 & -0.132 & -0.139 \\
re75         & -0.287 & \multicolumn{1}{c}{Contin.} & 0.042  & -0.007 & -0.036 & -0.039 & -0.049 \\\hline
\end{tabular}

\caption{Covariate balance table before and after matching for job training example. 10, 7, 5, 3 and 1 are the indices of matches. The other numbers in the table are the standardized difference in means in the matched treated and control groups.
}
\label{table:balance_lalonde1}
\end{table}  




\begin{table}[h!]
\begin{threeparttable}
\centering
\begin{tabular}{llr}
\hline
Matching Index          & Number of Matched Units & Percentage of Treated Units Matched\\ \hline
1 & 185                    & 100.0\%                     \\
3 & 182                      &     98.4\%                  \\
5 & 176                      & 95.1\%                      \\
7  & 155                      & 83.8\%                     \\
10  & 116                      &  62.7\%                    \\ \hline
\end{tabular} 

\end{threeparttable}
\caption{Match sizes in the job training example.}
\label{table:percentage}
\end{table}  



Sometimes it can be easier to understand tradeoffs by using plots than by using tables.  In Figure \ref{fig:eval_graphs_lalonde1} we visualize tradeoffs in each distinct pairing of our three design goals (within-pair distance, marginal balance for \textcolor{black}{race}, and number of subjects excluded) using the \texttt{visualize} command.  Each of the plots shown here appears to show a roughly linear relationship between the two design goals in question at least for a subset of the tradeoff space (although tradeoff curves are not guaranteed to be linear in general, merely monotonic). Thinking about these lines can be helpful in getting an overall sense for the ``price" of keeping one more treated unit in terms of marginal balance, and it also helps suggest areas of the tradeoff space that might need to be explored more.  
\textcolor{black}{For example, if the researcher wanted to keep more than 70\% of the data but did not view the balance achieved in match 6 as desirable enough, the tradeoff curve suggests where additional matches between matches 6 and 4 might lie and what balance they might offer.  The additional matches could then be generated by another call to \texttt{dist\_bal\_match}, specifying penalty parameters between the values used to create matches 4 and 6.}


\begin{lstlisting}[language=R]
visualize(
          match_result, 
          x_axis = "exclude", 
          y_axis ="marginal"
         )
visualize(
          match_result,  
          x_axis = "exclude", 
          y_axis = "pair"
         )
visualize(
          match_result, 
          x_axis = "pair", 
          y_axis = "marginal"
         )

\end{lstlisting}

\subsubsection{\textcolor{black}{Choosing penalty parameters}}
\label{subsubsec:penalties}

\textcolor{black}{A key challenge in getting the most out of \texttt{dist\_bal\_match} is identifying a set of penalty parameters that will highlight a diverse and interesting range of matched solutions.  Earlier in Section \ref{subsec:distbal} we presented a set of such penalties in our initial example for purposes of exposition, but in practice an iterative process is required to settle on an informative set of penalties. As equation (\ref{eqn:penalized}) suggests, both the exclusion penalty $\rho_1$ and the marginal balance penalty $\rho_2$ may be interpreted on the scale of the third objective function $f_3$, which is given here by the average pairwise distance.  Thus $\rho_1$ gives the largest increase in average pairwise distance we are willing to accept in order to retain one more treated unit, and $\rho_2$ gives the largest increase in pairwise distance we would accept in order to improve the imbalance score by a single unit (or equivalently to improve the total variation distance by 1/$(2|\mathcal{M}|)$ as discussed in Section \ref{subsec:margbal}).  While these interpretations can be helpful, they typically do not obviate the need for manual exploration of the results obtained under different penalty combinations.}

\textcolor{black}{One tool that \texttt{MultiObjMatch} provides to help users quickly identify the most appropriate range of penalties to explore is an automated grid search.
 The automatic search starts by setting the balance penalty to 1 and choosing a range of values for the exclusion penalty based on the distribution of treated-control distances: the mean distance, the 25th and 75th percentiles of the distance distribution, the minimum and maximum distance, half the minimum distance and a large multiple of the maximum distance (the multiplier can be specified by the optional argument \texttt{rho\_max\_factor}), and a very small number equal to the numeric tolerance used by the algorithm (specified in optional argument \texttt{tol}).  After computing a match for each combination of these penalties, the search procedure identifies the matches that achieved maximal and minimal values for each of the penalized objective functions and extracts the associated penalty combinations $(\rho_1, \rho_2)$. Letting $\rho_1^{\min},\rho_1^{\max}$ and $\rho_2^{\min},\rho_2^{\max}$ be the minimal and maximal penalties extracted for the exclusion and balance functions respectively, the algorithm explores a grid of matches in the space $[0.1\rho_1^{\min}, 10\rho_1^{\max}] \times [0.1\rho_2^{\min}, 10\rho_2^{\max}]$.
 This process then iterates, with the number of iterations controlled by the \texttt{max\_iter} parameter (equal to 1 by default).}


\textcolor{black}{
In practice, we recommend starting with the automated search process, examining the resulting matches and their associated $\rho$-values, and then handpicking penalty combinations, both to augment the original set of matches with new penalty combinations to further explore interesting parts of the tradeoff space and to eliminate uninteresting or repetitive solutions in other parts of the space. 
This could include eliminating some penalties from consideration, choosing larger or smaller penalty values (if the most extreme matches constructed are not yet too extreme on one or more dimensions), or choosing penalty combinations between values already attempted to obtain finer resolution among matches in an interesting region. 
 It may also be necessary to iterate before settling on a final penalty vector. 
}

\begin{figure}[h!]
     \centering
     \begin{subfigure}[b]{0.3\textwidth}
         \centering
         \includegraphics[width=\textwidth, scale = 1.7]{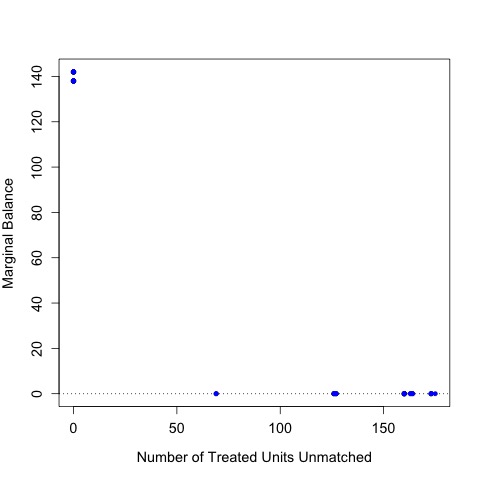}
         \label{fig:tv_unmatch1}
         \caption{}
     \end{subfigure}
     \hfill
     \begin{subfigure}[b]{0.3\textwidth}
         \centering
         \includegraphics[width=\textwidth, scale = 1.7]{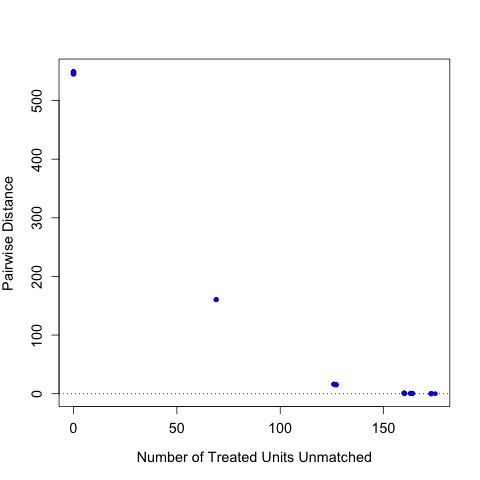}
         \label{fig:pairdist_unmatch1}
         \caption{}
     \end{subfigure}
     \hfill
     \begin{subfigure}[b]{0.3\textwidth}
         \centering
         \includegraphics[width=\textwidth, scale = 1.7]{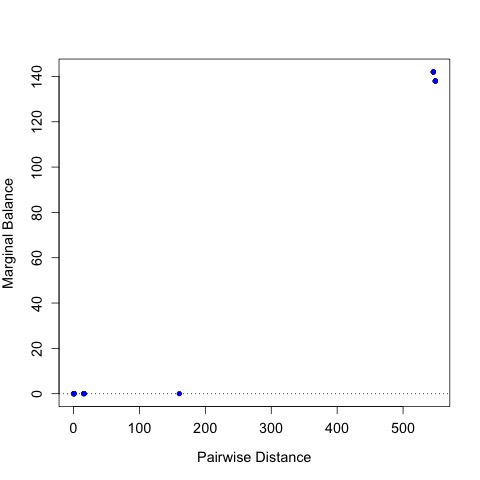}
         \label{fig:pairdist_tv1}
         \caption{}
     \end{subfigure}
        \caption{Objective values from initial grid search in job training example}
        \label{fig:param_search_visualization}
\end{figure}

\textcolor{black}{
In the job training example we follow this recipe, first running the automatic grid search and visualizing the results,  then successively tailoring the penalty vectors to produce the values used above.  Here \texttt{max\_iter} is initially left equal to its default value of 1 (in contrast to the the code in Section \ref{subsec:distbal} which specifies \texttt{max\_iter = 0} to suppress the grid search}).
\begin{lstlisting}[language=R]
initial_grid_search <- dist_bal_match(
  lalonde, 
  treat_col = treat_val, 
  marg_bal_col = bal_val, 
  dist_col = covariates
)    
\end{lstlisting}
\textcolor{black}{
The full set of solutions generated by the  initial grid search is shown in Figure \ref{fig:param_search_visualization}.
Tradeoffs are evident between the total pairwise distance and the number of treated units retained, but the marginal balance is either perfect or very poor (Figure \ref{fig:param_search_visualization}(a)). 
This suggests that it would be useful to test new, less extreme values of the balance penalty.}

\begin{figure}[H]
    \centering
    \includegraphics[width=\textwidth]{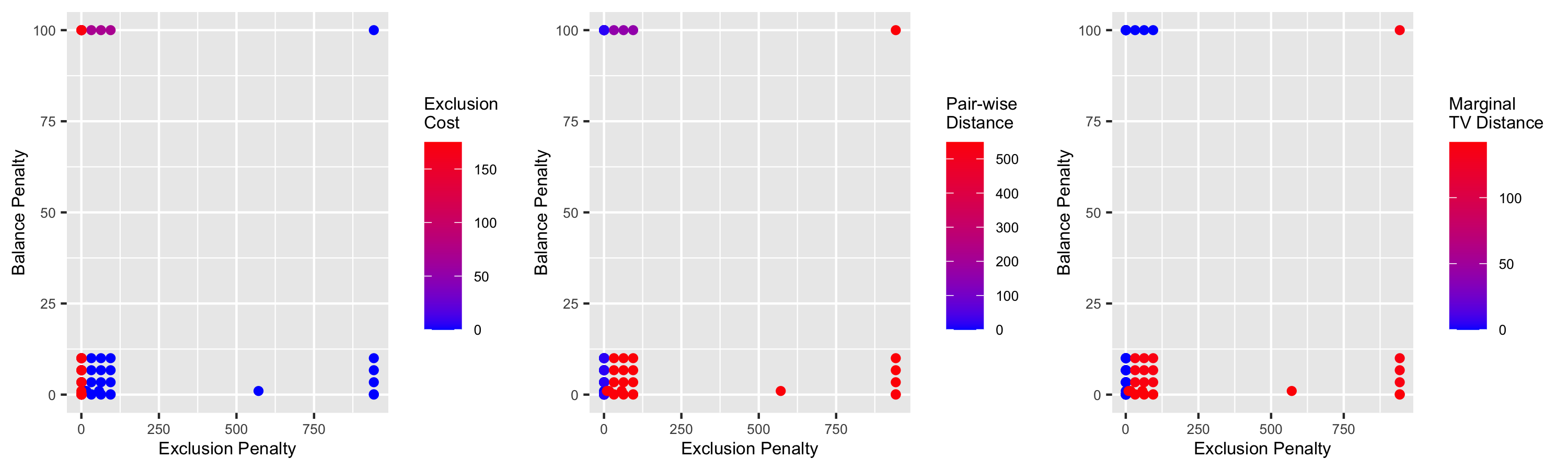}
    \caption{Objective function values and penalty values for initial grid search in job training example.}
    \label{fig:initial}
\end{figure}

\begin{figure}[H]
    \centering
    \includegraphics[width=\textwidth]{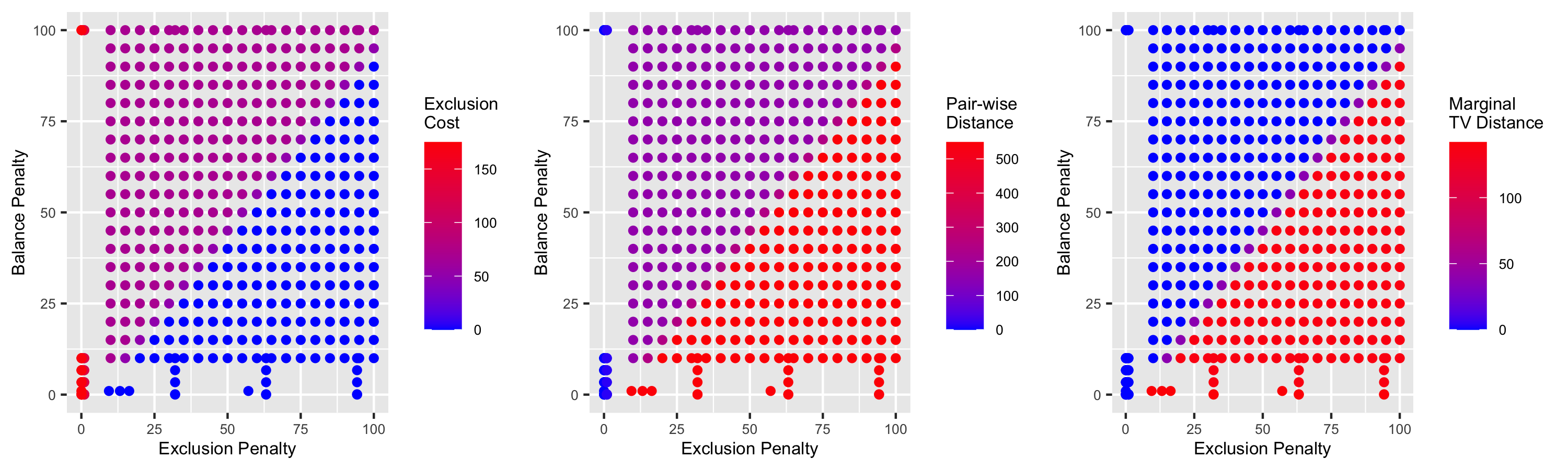}
    \caption{Objective function values over the 2-dimensional space of penalties for marginal total variation penalties between 0 and 100.}
    \label{fig:intermediate}
\end{figure}

\textcolor{black}{Figure \ref{fig:initial} facilitates this process by showing how objective costs vary with penalty values and showing where the gaps lie.  These plots are generated using the same \texttt{visualize} command introduced in Section \ref{subsec:display} but adding an additional \texttt{z\_axis} argument (code is shown below for the plot in the first panel of Figure \ref{fig:initial}).  We also specify \texttt{display\_index=FALSE} to suppress the printing of match indices on the plot, and provide an informative name for the color scale using the \texttt{zlab} argument.}
\begin{lstlisting}[language=R]
visualize(
           initial_grid_search, 
           x_axis = "exclusion_penalty", 
           y_axis = "balance_penalty", 
           z_axis = "exclude",
           display_index=FALSE, 
           zlab="Exclusion Cost"
          )
\end{lstlisting}          
\textcolor{black}{
Large matches occur in this plot only when the balance penalty is set to values below 100, so a natural next step is to explore the values in this range. 
We specify a grid of balance penalties from $0$ to $100$ while keeping the same grid of exclusion penalties as in the initial search. Figure \ref{fig:intermediate} reveals the results of this second attempt, in combination with the results from the initial grid search (the  \texttt{combine\_match\_result} command is used to concatenate results from the two \texttt{dist\_bal\_match} calls before \texttt{visualize} is called).}
\begin{lstlisting}[language=R]
match_result_handpick <- dist_bal_match(
  lalonde, 
  treat_col = treat_val, 
  marg_bal_col = bal_val, 
  exclusion_penalty = seq(10,100,5), 
  balance_penalty = seq(10,100,5), 
  dist_col = covariates, 
)
combined_results <- combine_match_result(
  match_result_handpick, 
  initial_grid_search
 )
\end{lstlisting}

\textcolor{black}{Now we see more nuanced choices available between extremes as exclusion penalties increase from $0$ to $100$. For each exclusion penalty value, there is clearly a small band of marginal penalty values where objective function values change meaningfully; beyond this small band, there is not much interesting variation across the objective functions. 
For our final examination of tradeoffs, this suggests fixing a single value for the exclusion penalty from the range shown in Figure \ref{fig:intermediate} (94.27, which is a value proposed by the auto-grid search and is the maximum value of the largest pair-wise robust Mahalanobis distance), and exclusion penalties in the range of 85 to 91, and these are the values we study in depth above in Section \ref{subsec:distbal}. 
}





\subsection{Preparing for outcome analysis}
After analyzing numerical and graphical diagnostics, the researcher can choose a match that strikes an appropriate balance of success on the specified design goals. 
In the tradeoff analysis for the job training example in Section \ref{subsec:distbal}, for instance, match \#5 reduces the imbalance on the Black variable to less than 0.2 standard deviations while still keeping 80\% of the sample and achieving excellent within-pair distance. 
Once a match has been selected, standard estimation and inference tools can be used to compare outcomes across matched groups \citep{rubin1979using, rosenbaum2002covariance, fogarty2020studentized}.  We demonstrate such an outcome analysis in Section \ref{subsec:outcomes}.

In many cases, it may be difficult to choose a single match from among a group of very similar matches with attractive properties. In these settings, we recommend taking all acceptable matches and running the analysis for each to confirm that results are stable across matches. This type of stability check may be formalized through devices such as the hacking interval \citep{morucci2018hypothesis, coker2021theory} and the PCS framework \citep{dwivedi2020stable}. Note that other forms of stability analysis are also a good idea to establish the empirical value of a matched analysis. In Section \ref{subsec:sens} we illustrate one such approach, sensitivity analysis for unobserved confounding \citep{rosenbaum2005sensitivity}. 

To extract the data needed for outcome analysis, the function \texttt{matched\_data} can be used; the researcher must supply the object produced by the original tradeoff command and the index of the match to be extracted. The extracted data contains the subset of \textcolor{black}{observations in the original data frame} that are \textcolor{black}{included in the match}, along with a matched pair indicator in the column \texttt{matchID}.
\begin{lstlisting}[language=R]
matched_data(match_result, 5)
\end{lstlisting}

\begin{table}[h!]
\centering
\textcolor{black}{\begin{tabular}{l*{5}{D{.}{.}{-1}}}
\hline
                & \multicolumn{1}{c}{10}     & \multicolumn{1}{c}{7}      & \multicolumn{1}{c}{5}      & \multicolumn{1}{c}{3}      & \multicolumn{1}{c}{1\tnote{a}} \\
\hline
age          & 0.149  & 0.074  & 0.042  & 0.035  & 0 \\
married      & -0.044 & -0.030  & -0.010  & -0.003 & 0 \\
educ         & 0.038  & 0.008  & 0.016  & 0.005  & 0 \\
nodegree     & 0.036  & 0.037  & 0.015  & 0.005  & 0 \\
race\_black  & 0.093  & 0.030   & 0.008  & 0.003  & 0 \\
race\_hispan & -0.035 & -0.012 & -0.003 & -0.001 & 0 \\
race\_white  & -0.058 & -0.019 & -0.005 & -0.002 & 0 \\
re74         & -0.134 & -0.057 & -0.022 & -0.007 & 0 \\
re75         & -0.109 & -0.041 & -0.024 & -0.008 & 0 \\
\hline
\end{tabular}}


\caption{\textcolor{black}{Covariate balance between matched treated units and all treated units,  for job training example. SMDs are calculated with the standard deviation for all treated units in the denominator.}}   
\label{table:representative_lalonde}
\end{table}

\section{Diabetic kidney disease and hospitalization costs}    
\label{sec:dkd}

We now use \texttt{MultiObjMatch} to construct a matched design for measuring the impact of diabetic kidney diseases (DKD) on hospitalization \textcolor{black}{cost}. 
The data set comes from the First Affiliated Hospital to Chongqing Medical University, China, a 3A hospital. Patients' hospitalization cost records from 2006-2016 are included in the data. There are 15,168 observations in the dataset, 
each including measurements for patient age, duration of the stay at the hospital in days, diagnosis and complications, and hospitalization cost (in Chinese Yuan). 
There are 3,463 subjects with DKD, whom we will denote as treated, and 11,705 without DKD whom we will denote as controls.

As in many other biomedical settings, the treatment cannot be randomly assigned due to practical (and ethical) constraints. Moreover, patients in the treatment group differ from those in the control group on important covariates before treatment. 
DKD patients are older and experience \textcolor{black}{a higher} incidence of microvascular complications. 
We will attempt to address these discrepancies by matching each patient with DKD to a patient without DKD with similar demographic characteristics and medical history. 

\subsection{Design Goals}
  
The primary goal is to form pairs of patients that are similar on all of a set of variables considered to be predictive of hospitalization cost (age, the presence of microvascular diseases, year of hospitalization, the presence of diabetic retinopathy (DR), and the presence of diabetic peripheral neuropathy (DPN)\textcolor{black}{)}, as measured by Mahalanobis distance. Another goal is to retain as many \textcolor{black}{DKD patients} as possible.  The three goals above are competing -- for example, pairwise closeness and marginal balance will be lower when all treated units are retained than when some treated units are excluded.

\begin{table}[h!]
\begin{threeparttable}
\centering
\begin{tabular}{llr}
\hline
              & type    & Standardized Differences   \\ \hline
age           & Contin. &0.589         \\
year\_2006    & Binary  & -0.051        \\
year\_2007    & Binary  & -0.009       \\
year\_2008    & Binary  & -0.106      \\
year\_2009    & Binary  & -0.037          \\
year\_2010    & Binary  & -0.051\\
year\_2011    & Binary  & 0.057\\
year\_2012    & Binary  & 0.001 \\
year\_2013    & Binary  & -0.031\\
year\_2014    & Binary  & 0.043 \\
year\_2015    & Binary  & 0.070 \\
year\_2016    & Binary  & 0.054 \\
sex           & Binary  & 0.076  \\
Number Micro & Contin. &1.631\\
DR        & Binary  & 0.414\\
DPN          & Binary  & 0.260  \\ \hline
\end{tabular}

\end{threeparttable}
\caption{Standardized differences in means of covariates \textcolor{black}{after initial} match \textcolor{black}{for DKD data. ``Number Micro" refers to the number of diagnosed micro-vascular complications, DR is diabetic retinopathy, and DPN is diabetic peripheral neuropathy.}}
\label{table:match_intial_attempt}
\end{table}

\textcolor{black}{ Researchers also desire marginal balance in the distribution of the number of microvascular complications, since 
 patients with more complications tend to receive a much higher medical bill, and since the strong association between microvascular complications and diabetic kidney diseases makes balance difficult to achieve. Even after an initial match on} pair-wise Euclidean distance without special attention to the marginal balance in the number of microvascular complications, the treated patients tend to have a much higher number of microvascular complications: \textcolor{black}{the} standardized difference in means \textcolor{black}{is} 1.631 \textcolor{black}{(Table \ref{table:match_intial_attempt}).  Enforcing some level marginal balance on microvascular complications will provide a more direct and effective way to remove this imbalance.}  

Finally, since the price level and other economic factors change year by year, matching exactly by year is important to keep outcomes comparable within matched sets. 

The primary study goal is to answer whether DKD increases hospitalization costs. There is a very strong correlation between total hospitalization cost and the number of days of hospitalization, as previous researchers have identified\citep{cao2015factors, ward2014direct}. To avoid the imbalance of the distribution of the duration of hospitalization stays, which may further affect the total cost, daily hospitalization cost, the total cost divided by the total days in the hospital, is used as the outcome of interest. 


\subsection{Matching Result}
\textcolor{black}{We use \texttt{dist\_bal\_match} to investigate the tradeoff.  As in Section \ref{subsubsec:penalties} we begin with an automatic grid search and refine the set of solutions presented; for full details (and specific values for the \texttt{exclusion\_penalty} and \texttt{balance\_penalty} arguments) see the Appendix (Section \ref{sec:dkd_hyperparam_searching}).}  
Since this match is much larger than the one conducted in the job training example, we impose a propensity caliper forbidding matches between units more than 0.25 standard deviations apart in their estimated propensity scores; in combination  with our choice to match exactly on year, this reduces the number of possible matches considered and makes the problem more computationally tractable \citep{pimentel2015large}. The \texttt{propensity\_col = covariates} argument specifies the columns on which to fit the propensity score, the \texttt{caliper\_option} argument gives the size of the caliper, and the \texttt{exact} argument specifies our exact matching constraint.
\begin{lstlisting}[language=R]
dist_bal_match(
  dkd, 
  treat_col = treat_val,
  marg_bal_col = bal_val, 
  dist_col  = covariates,
  exact_col = "year", 
  exclusion_penalty = rho_exclusion,
  balance_penalty = rho_balance,
  propensity_col = covariates,
  caliper_option = 0.25,
  max_iter = 0
)
\end{lstlisting}

\begin{figure}[h!]
     \centering
     \begin{subfigure}[b]{0.45\textwidth}
         \centering
         \includegraphics[width=\textwidth, scale = 1.7]{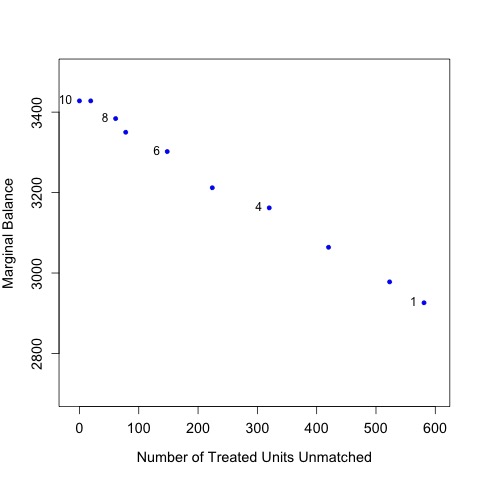}
         \caption{}
     \end{subfigure}
     \hfill
     \begin{subfigure}[b]{0.45\textwidth}
         \centering
         \includegraphics[width=\textwidth, scale = 1.7]{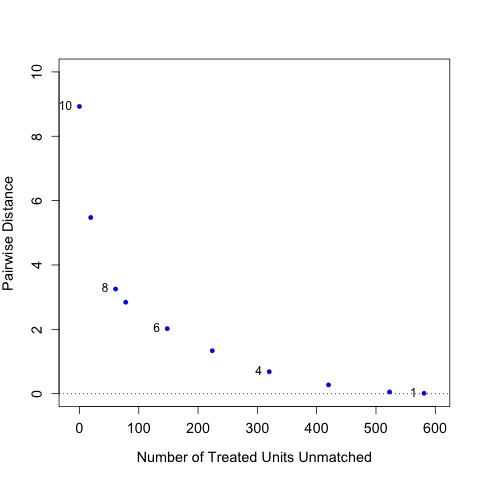}
         \caption{}
     \end{subfigure}
     \hfill
     \begin{subfigure}[b]{0.45\textwidth}
         \centering
         \includegraphics[width=\textwidth, scale = 1.7]{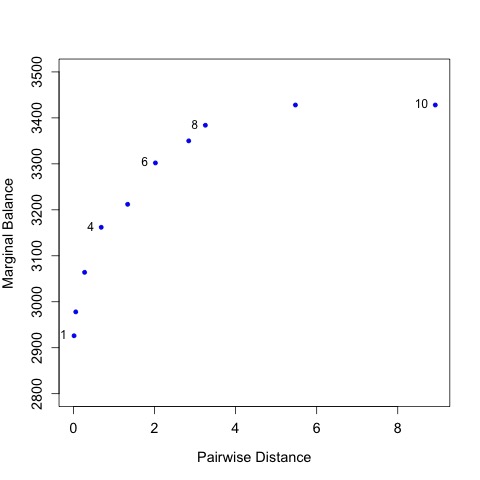}
         \caption{}
     \end{subfigure}
        \caption{Three graphs that visualize the trade-offs among three objectives}
        \label{fig:eval_graphs_dkd1}
\end{figure}

Tables \ref{table:balance_dkd1}-\ref{table:percentage_dkd} and Figure \ref{fig:eval_graphs_dkd1} depict the tradeoffs between three design goals for the DKD study. \textcolor{black}{While \texttt{summary} prints information for all matches computed, in the interest of concision, we display only a subset of this information,  for five interesting matches.} Covariate balance is improved after matching across all the Pareto optimal matches. 
After matching, even though \textcolor{black}{imbalance in the number of microvascular complications remains worse than for the other covariates, it is much better than in the naive match of Table \ref{table:match_intial_attempt}.} 

One important takeaway from \textcolor{black}{Table \ref{table:percentage_dkd} and Figure \ref{fig:eval_graphs_dkd1}} is that sacrificing a few patients and accepting a smaller sample size may confer substantial improvements in the other objectives. Notice that while match \#8 drops \textcolor{black}{61} more units than match \#10  (\textcolor{black}{1.8\%} of the treated group), the pairwise distance sum for match \#8 is close to half of that in match \#10. Moreover,  \textcolor{black}{matches with even smaller sample sizes} achieve a close-to-zero pair-wise distance sum. On balance it seems worth it to drop at least 100 treated subjects or so (as match \#6 does), and researchers may consider dropping more to further improve the quality of the match, perhaps by instead choosing match \#4 or one of its neighbors.  On the other hand match \#1, which drops over 1/3 of the treated subjects, is likely too extreme in its readiness to exclude subjects to be the best choice. 


\begin{table}[h!]
\begin{threeparttable}
\centering
\begin{tabular}{llllllll}
\hline
              & type    & Before Matching & 10     & 8      & 6      & 4      & 1      \\ \hline
age           & Contin. & 0.513           & -0.064 & -0.078 & -0.060 & -0.061 & -0.072 \\
sex           & Binary  & 0.038           & -0.016  & -0.004  & -0.019  & -0.013  & -0.008  \\
Number Micro & Contin. & 1.789           & -0.112 & -0.108 & -0.107 & -0.094 & -0.106 \\
DR           & Binary  & 0.197           & -0.018 & -0.017 & -0.016 & -0.012 & -0.014 \\
DPN          & Binary  & 0.129           & -0.025 & -0.021 & -0.022 & -0.019 & -0.027 \\
\hline
\end{tabular}

\end{threeparttable}
\caption{Standardized differences in means for covariates before and after matching \textcolor{black}{across five of the matches produced by \texttt{dist\_bal\_match} (each matching exactly on year).  The numbers in the top row are indices for individual matches. As in \ref{table:match_intial_attempt}, ``Number Micro" refers to the number of diagnosed micro-vascular complications, DR is diabetic retinopathy, and DPN is diabetic peripheral neuropathy.}}
\label{table:balance_dkd1}
\end{table}

\begin{table}[h!]
\begin{threeparttable}
\centering
\begin{tabular}{rrr}
\hline
Matching Index & Number of Matched Units & Percentage of Treated Units Matched \\ \hline
10             & 3463                    & 100.0\%                       \\
8              & 3402                    & 98.2\%                       \\
6              & 3315                    & 95.7\%                       \\
4              & 3143                    & 90.7\%                       \\
1              & 2882                    & 83.2\%                       \\ \hline
\end{tabular}
\end{threeparttable}
\caption{Number and percentage of matched units for \textcolor{black}{five of the} matches \textcolor{black}{produced by \texttt{dist\_bal\_match} in the} DKD example.}
\label{table:percentage_dkd}
\end{table}

\subsection{Outcome Analysis}   
\label{subsec:outcomes}

After matching, we can analyze DKD's effect outcome by randomization inference, permuting treatment assignments within matched pairs and recomputing the difference-in-means statistic to create a null distribution.  Since treatments were not assigned randomly, this procedure is technically a quasi-randomization test in the language of \citet{zhang2023randomization}. Point estimates, p-values, and 95\% confidence intervals obtained by inverting the randomization test \citep{rosenbaum2015two} are reported in Table \ref{table:outcome}. \textcolor{black}{All analyses show the impact of DKD to be a significant increase in daily cost.}



\begin{table}[h!]
\begin{threeparttable}
\centering
\begin{tabular}{r|rrrr}
\hline
Match Index & Point Estimate & 95\% Confidence Interval & P-value & $\Gamma$ \\ \hline 
10 & 49.7 & [14.1, 81.1] & 0.001 & 2.27\\
8 & 47.6 &[12.1, 76.0] & 0.002 & 2.15 \\
6 & 55.1 &[23.7, 84.7] & 0.000  & 2.20 \\
4 & 48.9 &[18.7, 79.3] & 0.002 & 2.20 \\
1 & 52.3 & [19.6, 78.8] & 0.002 & 2.00 \\\hline
\end{tabular}
\end{threeparttable}
\caption{Randomization inference and sensitivity analysis for the five matches.  $\Gamma$ measures the minimum degree of unobserved confounding (where true treatment odds are allowed to differ for apparently identical subjects by a factor of up to $\Gamma$)
needed to explain the observed data in the absence of an effect.}
\label{table:outcome}
\end{table}  

\subsection{Sensitivity Analysis}
\label{subsec:sens}
Sensitivity analysis aims to quantify how strong unmeasured confounders would need to be in order to explain observed patterns in the data in the absence of a treatment effect. Although observed covariates judged to be predictive of cost for DKD patients are all included for matching, not all relevant background and medical history \textcolor{black}{information is} recorded in the available patient data.

\textcolor{black}{We  conduct} sensitivity analysis for matches \#10, \#8, \#6, \#4, and \#1 \textcolor{black}{under a model that allows an adversarial unobserved confounder to influence relative treatment odds within each matched pair up to a fixed multiplicative factor $\Gamma$.} Technical details can be found in \citep{rosenbaum2005sensitivity} and \citep{rosenbaum2006package}. In Table \ref{table:outcome}, the largest $\Gamma$ that yields a one-sided p-value smaller or equal to 0.05 are reported for the five matches. 

The $\Gamma$ values for the five matches are all slightly above 2. This means that an unobserved confounder leading patients that look identical (on observed covariates) to differ in their odds of DKD by a factor of 2 could explain the observed data, but that unobserved confounders leading to a difference less than 2 could not. 
As such, unobserved confounders with only weak or tiny links to DKD are insufficient to cast doubt on the observed results; any meaningful confounder suggested by a critic of our study must demonstrate at least a moderately strong connection to DKD, or
the conclusion that DKD drives higher hospitalization costs would remain unaltered.

\section{Discussion and Opportunities}\label{sec:discussion_opportunities}
\subsection{MultiObjMatch enhances choice among matched designs}

\texttt{MultiObjMatch} produces matches that satisfy three design goals with varying degree, allowing for greater flexibility than 
matching with single or even double objectives. As seen in the examples in Sections \ref{sec:lalonde},  \ref{sec:dkd}, \textcolor{black}{and \ref{subsec:twodist}}, objectives can include multiple distance metrics, marginal balance measures, and sample size.  
Not all possible tradeoffs are ultimately interesting --- sometimes multiple design goals can all be achieved reasonably well --- but most of the time, especially when reducing sample size is an option, tradeoff analysis reveals matches too extreme in both directions as well as a more attractive group of matches in the middle that place non-trivial emphasis on all goals.  In both case studies, for example, excluding some number of treatment units ultimately proved beneficial for achieving balance on a key variable.  Visualization tools are especially useful for gaining an understanding of the relative costs and benefits of different candidate matches.  Choosing a single match from a family of similar matches can be difficult, and stability analysis, under which outcome analysis is conducted and reported for all plausible matches, is recommended in practice.

\subsection{Computational cost}  
\texttt{MultiObjMatch} solves individual matching problems by representing them internally as minimum-cost network flow problems. 
Minimum-cost network flow is a canonical problem in operations research for which efficient general-purpose solvers are available. \textcolor{black}{The LEMON open source graph library implements several solution algorithms \citep{dezsHo2011lemon}. One such algorithm based on a strategy called cycle cancelling \citep{klein1967primal, goldberg1989finding} is used via the \texttt{rlemon} package \citep{agarwal2023rlemon} as the default network-flow solver by leading optimal matching packages such as \texttt{optmatch} and \texttt{rcbalance}. 
 \texttt{MultiObjMatch} follows their lead in using this well-validated solver.}

\textcolor{black}{Despite the use of a state-of-the art efficient solver from a library to handle minimum-cost network flow computation, solving the network flow problem iteratively remains} the most computationally expensive component of \texttt{MultiObjMatch}'s operations. Within each iteration, a different combination of penalty coefficients is passed in for the objective functions and the problem must be solved again. Two factors that heavily influence worst-case bounds on computational speed for the network flow problem are the number of treated-control pairings considered and the maximum cost in the network; as documented in \citet{pimentel2019optimal}, these quantities appear to have substantial influence on practical performance as well. More restrictive calipers and more strict exact matching criteria leave the network structure more sparse, and therefore, faster to be solved. However, when the network is too sparse, the algorithm might be restricted in its ability to explore all interesting matches; for instance, tight calipers often make it impossible to match all treated units, requiring some to be excluded in all cases. Researchers may want to start with high levels of sparsity and gradually reduce sparsity in order to relax such restrictions without causing computation to choke.

\subsection{Directions for future work}

We have focused on examining tradeoffs between two objectives at a time in Section 3 and Section 4, holding the third fixed at either a very low or a very high level of priority. In principle, more complex higher-dimensional tradeoffs may be scientifically interesting, but efficient exploration algorithms to describe the full set of Pareto optimal matches with more than two objective functions have not been clearly described.  Although future research into such algorithms is needed, the functionality provided by \texttt{MultiObjMatch} may be helpful as a building block for such work.

Two possibilities for future computational improvement are also worth mentioning.  First, the process of solving network-flow problems for different combinations of penalty coefficients is in principle highly parallelizable.  While the current implementation of \texttt{MultiObjMatch} solves network flow problems specified by a list of penalty parameters iteratively, a parallel approach could solve all the instances specified by a vector of tuning parameters simultaneously.  Such an approach could potentially be implemented by leveraging existing parallel computing frameworks in R, such as \citet{schmidberger2009state, denwood2016runjags, suzuki2006pvclust}.  While it might still be necessary to conduct several batches of solution operations, this would surely improve performance.  The other potential improvement is using the solution to one network flow problem to create a ``warm start" for the next network problem, enabling quicker convergence to an optimal solution \citep{yildirim2002warm}.  This would require modification of the inputs to the network-flow problem.  Unfortunately, the two possible improvements would not seem to be mutually compatible, since parallel runs could not easily take advantage of each other's solutions as warm starts.


\section*{Acknowledgements}

Thanks to Zhihong Wang, Qifu Li, Jinbo Hu, and Xiangjun Chen from the Department of Endocrinology, the First Affiliated Hospital to Chongqing Medical University for data access and biomedical professsonal insights on the diabetic kidney disease case study. Thanks to Amanda Glazer (UC Berkeley) and Shiyuan Li (Peking University) and Chutong Gao (Northwestern University) for suggestions on R package development and early drafts.

\bibliographystyle{asa}
\bibliography{ref}

\begin{thebibliography}{72}
\newcommand{\enquote}[1]{``#1''}
\expandafter\ifx\csname natexlab\endcsname\relax\def\natexlab#1{#1}\fi

\bibitem[{Agarwal et~al.(2022)Agarwal, Tewari, and Errickson}]{agarwal2023rlemon}
Agarwal, A., Tewari, A., and Errickson, J. (2022), \textit{rlemon: R Access to LEMON Graph Algorithms}, r package version 0.1.0.

\bibitem[{Antonelli et~al.(2018)Antonelli, Cefalu, Palmer, and Agniel}]{antonelli2018doubly}
Antonelli, J., Cefalu, M., Palmer, N., and Agniel, D. (2018), \enquote{Doubly robust matching estimators for high dimensional confounding adjustment,} \textit{Biometrics}, 74, 1171--1179.

\bibitem[{Austin(2008)}]{austin2008critical}
Austin, P.~C. (2008), \enquote{A critical appraisal of propensity-score matching in the medical literature between 1996 and 2003,} \textit{Statistics in medicine}, 27, 2037--2049.

\bibitem[{Bennett et~al.(2020)Bennett, Vielma, and Zubizarreta}]{bennett2020building}
Bennett, M., Vielma, J.~P., and Zubizarreta, J.~R. (2020), \enquote{Building representative matched samples with multi-valued treatments in large observational studies,} \textit{Journal of computational and graphical statistics}, 29, 744--757.

\bibitem[{Cai et~al.(2018)Cai, Li, Cui, You, and Golden}]{cai2018trends}
Cai, L., Li, X., Cui, W., You, D., and Golden, A.~R. (2018), \enquote{Trends in diabetes and pre-diabetes prevalence and diabetes awareness, treatment and control across socioeconomic gradients in rural southwest China,} \textit{Journal of Public Health}, 40, 375--380.

\bibitem[{Caliendo and Kopeinig(2008)}]{caliendo2008some}
Caliendo, M. and Kopeinig, S. (2008), \enquote{Some practical guidance for the implementation of propensity score matching,} \textit{Journal of economic surveys}, 22, 31--72.

\bibitem[{Cao et~al.(2015)Cao, Wang, Zhang, Zhao, and Li}]{cao2015factors}
Cao, P., Wang, K., Zhang, H., Zhao, R., and Li, C. (2015), \enquote{Factors influencing the hospitalization costs of patients with type 2 diabetes,} \textit{Asia Pacific Journal of Public Health}, 27, 55S--60S.

\bibitem[{Coker et~al.(2021)Coker, Rudin, and King}]{coker2021theory}
Coker, B., Rudin, C., and King, G. (2021), \enquote{A theory of statistical inference for ensuring the robustness of scientific results,} \textit{Management Science}, 67, 6174--6197.

\bibitem[{de~los Angeles~Resa and Zubizarreta(2016)}]{resa2016evaluation}
de~los Angeles~Resa, M. and Zubizarreta, J.~R. (2016), \enquote{Evaluation of subset matching methods and forms of covariate balance,} \textit{Statistics in medicine}, 35, 4961--4979.

\bibitem[{Dehejia and Wahba(1999)}]{dehejia1999causal}
Dehejia, R.~H. and Wahba, S. (1999), \enquote{Causal effects in nonexperimental studies: Reevaluating the evaluation of training programs,} \textit{Journal of the American statistical Association}, 94, 1053--1062.

\bibitem[{Denwood et~al.(2016)}]{denwood2016runjags}
Denwood, M.~J. et~al. (2016), \enquote{runjags: An R package providing interface utilities, model templates, parallel computing methods and additional distributions for MCMC models in JAGS,} \textit{Journal of Statistical Software}, 71, 1--25.

\bibitem[{Dezs{\H{o}} et~al.(2011)Dezs{\H{o}}, J{\"u}ttner, and Kov{\'a}cs}]{dezsHo2011lemon}
Dezs{\H{o}}, B., J{\"u}ttner, A., and Kov{\'a}cs, P. (2011), \enquote{LEMON--an open source C++ graph template library,} \textit{Electronic notes in theoretical computer science}, 264, 23--45.

\bibitem[{Diamond and Sekhon(2013)}]{diamond2013genetic}
Diamond, A. and Sekhon, J.~S. (2013), \enquote{Genetic matching for estimating causal effects: A general multivariate matching method for achieving balance in observational studies,} \textit{Review of Economics and Statistics}, 95, 932--945.

\bibitem[{Dwivedi et~al.(2020)Dwivedi, Tan, Park, Wei, Horgan, Madigan, and Yu}]{dwivedi2020stable}
Dwivedi, R., Tan, Y.~S., Park, B., Wei, M., Horgan, K., Madigan, D., and Yu, B. (2020), \enquote{Stable discovery of interpretable subgroups via calibration in causal studies,} \textit{International Statistical Review}, 88, S135--S178.

\bibitem[{Farshchi et~al.(2014)Farshchi, Esteghamati, Sari, Kebriaeezadeh, Abdollahi, Dorkoosh, Khamseh, Aghili, Keshtkar, and Ebadi}]{farshchi2014cost}
Farshchi, A., Esteghamati, A., Sari, A.~A., Kebriaeezadeh, A., Abdollahi, M., Dorkoosh, F.~A., Khamseh, M.~E., Aghili, R., Keshtkar, A., and Ebadi, M. (2014), \enquote{The cost of diabetes chronic complications among Iranian people with type 2 diabetes mellitus,} \textit{Journal of Diabetes \& Metabolic Disorders}, 13, 42.

\bibitem[{Fogarty(2020)}]{fogarty2020studentized}
Fogarty, C.~B. (2020), \enquote{Studentized sensitivity analysis for the sample average treatment effect in paired observational studies,} \textit{Journal of the American Statistical Association}, 115, 1518--1530.

\bibitem[{Goldberg and Tarjan(1989)}]{goldberg1989finding}
Goldberg, A.~V. and Tarjan, R.~E. (1989), \enquote{Finding minimum-cost circulations by canceling negative cycles,} \textit{Journal of the ACM (JACM)}, 36, 873--886.

\bibitem[{Greifer(2016)}]{greifer2016covariate}
Greifer, N. (2016), \enquote{Covariate balance tables and plots: A guide to the cobalt package,} .

\bibitem[{Hansen(2004)}]{hansen2004full}
Hansen, B.~B. (2004), \enquote{Full matching in an observational study of coaching for the SAT,} \textit{Journal of the American Statistical Association}, 99, 609--618.

\bibitem[{Hansen(2007)}]{hansen2007optmatch}
--- (2007), \enquote{Optmatch: Flexible, optimal matching for observational studies,} \textit{New Functions for Multivariate Analysis}, 7, 18--24.

\bibitem[{Hansen(2008)}]{hansen2008prognostic}
--- (2008), \enquote{The prognostic analogue of the propensity score,} \textit{Biometrika}, 95, 481--488.

\bibitem[{Hansen et~al.(2019)Hansen, Fredrickson, Fredrickson, Rcpp, and Rcpp}]{hansen2019package}
Hansen, B.~B., Fredrickson, M., Fredrickson, M. M.~M., Rcpp, L., and Rcpp, I. (2019), \enquote{Package ‘optmatch’,} \textit{Available on https://cran. r-project. org/web/packages/optmatch/optmatch. pdf (last accessed on 10 October 2015)}.

\bibitem[{Ho et~al.(2007)Ho, Imai, King, and Stuart}]{ho2007matching}
Ho, D.~E., Imai, K., King, G., and Stuart, E.~A. (2007), \enquote{Matching as nonparametric preprocessing for reducing model dependence in parametric causal inference,} \textit{Political analysis}, 15, 199--236.

\bibitem[{Ho et~al.(2011)Ho, Imai, King, and Stuart}]{matchit}
--- (2011), \textit{MatchIt: Nonparametric Preprocessing for Parametric Causal Inference}, r package version 3.0.2.

\bibitem[{Iacus et~al.(2012{\natexlab{a}})Iacus, King, and Porro}]{iacus2012causal}
Iacus, S.~M., King, G., and Porro, G. (2012{\natexlab{a}}), \enquote{Causal inference without balance checking: Coarsened exact matching,} \textit{Political analysis}, 20, 1--24.

\bibitem[{Iacus et~al.(2012{\natexlab{b}})Iacus, King, and Porro}]{cem}
--- (2012{\natexlab{b}}), \textit{cem: Coarsened Exact Matching in Stata}, r package version 1.1.23.

\bibitem[{King et~al.(2016)King, Lucas, and Nielsen}]{king2016matchingfrontier}
King, G., Lucas, C., and Nielsen, R. (2016), \enquote{MatchingFrontier: Automated Matching for Causal Inference,} \textit{R package version}, 2.

\bibitem[{King et~al.(2017)King, Lucas, and Nielsen}]{king2017balance}
King, G., Lucas, C., and Nielsen, R.~A. (2017), \enquote{The balance-sample size frontier in matching methods for causal inference,} \textit{American Journal of Political Science}, 61, 473--489.

\bibitem[{King and Nielsen(2019)}]{king2019propensity}
King, G. and Nielsen, R. (2019), \enquote{Why propensity scores should not be used for matching,} \textit{Political analysis}, 27, 435--454.

\bibitem[{Klein(1967)}]{klein1967primal}
Klein, M. (1967), \enquote{A primal method for minimal cost flows with applications to the assignment and transportation problems,} \textit{Management Science}, 14, 205--220.

\bibitem[{LaLonde(1986)}]{lalonde1986evaluating}
LaLonde, R.~J. (1986), \enquote{Evaluating the econometric evaluations of training programs with experimental data,} \textit{The American economic review}, 604--620.

\bibitem[{Lee et~al.(2010)Lee, Lessler, and Stuart}]{lee2010improving}
Lee, B.~K., Lessler, J., and Stuart, E.~A. (2010), \enquote{Improving propensity score weighting using machine learning,} \textit{Statistics in medicine}, 29, 337--346.

\bibitem[{Liao et~al.(2023)Liao, Zhu, Ngo, Chehab, and Pimentel}]{liao2023using}
Liao, L.~D., Zhu, Y., Ngo, A.~L., Chehab, R.~F., and Pimentel, S.~D. (2023), \enquote{Using Joint Variable Importance Plots to Prioritize Variables in Assessing the Impact of Glyburide on Adverse Birth Outcomes,} \textit{arXiv preprint arXiv:2301.09754}.

\bibitem[{Ming and Rosenbaum(2000)}]{ming2000substantial}
Ming, K. and Rosenbaum, P.~R. (2000), \enquote{Substantial gains in bias reduction from matching with a variable number of controls,} \textit{Biometrics}, 56, 118--124.

\bibitem[{Morgan and Winship(2015)}]{morgan2015counterfactuals}
Morgan, S.~L. and Winship, C. (2015), \textit{Counterfactuals and causal inference}, Cambridge University Press.

\bibitem[{Morucci et~al.(2018)Morucci, Noor-E-Alam, and Rudin}]{morucci2018hypothesis}
Morucci, M., Noor-E-Alam, M., and Rudin, C. (2018), \enquote{Hypothesis tests that are robust to choice of matching method,} \textit{arXiv preprint arXiv:1812.02227}.

\bibitem[{Oakes and Johnson(2006)}]{oakes2006propensity}
Oakes, J.~M. and Johnson, P.~J. (2006), \enquote{Propensity score matching for social epidemiology,} \textit{Methods in social epidemiology}, 1, 370--393.

\bibitem[{Pimentel(2016)}]{pimentel2016large}
Pimentel, S.~D. (2016), \enquote{Large, sparse optimal matching with R package rcbalance,} \textit{Observational Studies}, 2, 4--23.

\bibitem[{Pimentel(2023)}]{pimentel2023fine}
--- (2023), \enquote{Fine Balance and Its Variations in Modern Optimal Matching,} in \textit{Handbook of Matching and Weighting Adjustments for Causal Inference}, Chapman and Hall/CRC, pp. 105--134.

\bibitem[{Pimentel and Kelz(2019)}]{pimentel2019optimal}
Pimentel, S.~D. and Kelz, R.~R. (2019), \enquote{Optimal Tradeoffs in Matched Designs Comparing US-Trained and Internationally Trained Surgeons,} \textit{Journal of the American Statistical Association}, 1--14.

\bibitem[{Pimentel et~al.(2015)Pimentel, Kelz, Silber, and Rosenbaum}]{pimentel2015large}
Pimentel, S.~D., Kelz, R.~R., Silber, J.~H., and Rosenbaum, P.~R. (2015), \enquote{Large, sparse optimal matching with refined covariate balance in an observational study of the health outcomes produced by new surgeons,} \textit{Journal of the American Statistical Association}, 110, 515--527.

\bibitem[{Pimentel et~al.(2016)Pimentel, Small, and Rosenbaum}]{pimentel2016constructed}
Pimentel, S.~D., Small, D.~S., and Rosenbaum, P.~R. (2016), \enquote{Constructed second control groups and attenuation of unmeasured biases,} \textit{Journal of the American Statistical Association}, 111, 1157--1167.

\bibitem[{Rosenbaum(1984)}]{rosenbaum1984consequences}
Rosenbaum, P.~R. (1984), \enquote{The consequences of adjustment for a concomitant variable that has been affected by the treatment,} \textit{Journal of the Royal Statistical Society: Series A (General)}, 147, 656--666.

\bibitem[{Rosenbaum(2002{\natexlab{a}})}]{rosenbaum2002covariance}
--- (2002{\natexlab{a}}), \enquote{Covariance adjustment in randomized experiments and observational studies,} \textit{Statistical Science}, 17, 286--327.

\bibitem[{Rosenbaum(2002{\natexlab{b}})}]{rosenbaum2002overt}
--- (2002{\natexlab{b}}), \enquote{Overt bias in observational studies,} in \textit{Observational studies}, Springer, pp. 71--104.

\bibitem[{Rosenbaum(2005{\natexlab{a}})}]{rosenbaum2005heterogeneity}
--- (2005{\natexlab{a}}), \enquote{Heterogeneity and causality: Unit heterogeneity and design sensitivity in observational studies,} \textit{The American Statistician}, 59, 147--152.

\bibitem[{Rosenbaum(2005{\natexlab{b}})}]{rosenbaum2005sensitivity}
--- (2005{\natexlab{b}}), \enquote{Sensitivity analysis in observational studies,} \textit{Encyclopedia of statistics in behavioral science}.

\bibitem[{Rosenbaum(2012)}]{rosenbaum2012optimal}
--- (2012), \enquote{Optimal matching of an optimally chosen subset in observational studies,} \textit{Journal of Computational and Graphical Statistics}, 21, 57--71.

\bibitem[{Rosenbaum(2015)}]{rosenbaum2015two}
--- (2015), \enquote{Two R packages for sensitivity analysis in observational studies,} \textit{Observational Studies}, 1, 1--17.

\bibitem[{Rosenbaum(2020)}]{rosenbaum2020design}
--- (2020), \textit{Design of observational studies}, Springer, 2nd ed.

\bibitem[{Rosenbaum and Rosenbaum(2006)}]{rosenbaum2006package}
Rosenbaum, P.~R. and Rosenbaum, M. P.~R. (2006), \enquote{Package ‘sensitivitymv’,} .

\bibitem[{Rosenbaum et~al.(2007)Rosenbaum, Ross, and Silber}]{rosenbaum2007minimum}
Rosenbaum, P.~R., Ross, R.~N., and Silber, J.~H. (2007), \enquote{Minimum distance matched sampling with fine balance in an observational study of treatment for ovarian cancer,} \textit{Journal of the American Statistical Association}, 102, 75--83.

\bibitem[{Rosenbaum and Rubin(1983)}]{rosenbaum1983central}
Rosenbaum, P.~R. and Rubin, D.~B. (1983), \enquote{The central role of the propensity score in observational studies for causal effects,} \textit{Biometrika}, 70, 41--55.

\bibitem[{Rosenbaum and Rubin(1985)}]{rosenbaum1985constructing}
--- (1985), \enquote{Constructing a control group using multivariate matched sampling methods that incorporate the propensity score,} \textit{The American Statistician}, 39, 33--38.

\bibitem[{Rosenbaum et~al.(2010)}]{rosenbaum2010design}
Rosenbaum, P.~R. et~al. (2010), \textit{Design of observational studies}, vol.~10, Springer.

\bibitem[{Rubin(1979)}]{rubin1979using}
Rubin, D.~B. (1979), \enquote{Using multivariate matched sampling and regression adjustment to control bias in observational studies,} \textit{Journal of the American Statistical Association}, 74, 318--328.

\bibitem[{Rubin(2001)}]{rubin2001using}
--- (2001), \enquote{Using propensity scores to help design observational studies: application to the tobacco litigation,} \textit{Health Services and Outcomes Research Methodology}, 2, 169--188.

\bibitem[{Rubin and Thomas(1996)}]{rubin1996matching}
Rubin, D.~B. and Thomas, N. (1996), \enquote{Matching using estimated propensity scores: relating theory to practice,} \textit{Biometrics}, 249--264.

\bibitem[{Rubin and Thomas(2000)}]{rubin2000combining}
--- (2000), \enquote{Combining propensity score matching with additional adjustments for prognostic covariates,} \textit{Journal of the American Statistical Association}, 95, 573--585.

\bibitem[{Schmidberger et~al.(2009)Schmidberger, Morgan, Eddelbuettel, Yu, Tierney, and Mansmann}]{schmidberger2009state}
Schmidberger, M., Morgan, M., Eddelbuettel, D., Yu, H., Tierney, L., and Mansmann, U. (2009), \enquote{State-of-the-art in Parallel Computing with R,} \textit{Journal of Statistical Software}, 47.

\bibitem[{Stuart(2010)}]{stuart2010matching}
Stuart, E.~A. (2010), \enquote{Matching methods for causal inference: A review and a look forward,} \textit{Statistical science: a review journal of the Institute of Mathematical Statistics}, 25, 1.

\bibitem[{Stuart et~al.(2013)Stuart, Lee, and Leacy}]{stuart2013prognostic}
Stuart, E.~A., Lee, B.~K., and Leacy, F.~P. (2013), \enquote{Prognostic score--based balance measures can be a useful diagnostic for propensity score methods in comparative effectiveness research,} \textit{Journal of clinical epidemiology}, 66, S84--S90.

\bibitem[{Suzuki and Shimodaira(2006)}]{suzuki2006pvclust}
Suzuki, R. and Shimodaira, H. (2006), \enquote{Pvclust: an R package for assessing the uncertainty in hierarchical clustering,} \textit{Bioinformatics}, 22, 1540--1542.

\bibitem[{Thuong(2020)}]{thuong2020impact}
Thuong, N. T.~T. (2020), \enquote{Impact of health insurance on healthcare utilisation patterns in Vietnam: a survey-based analysis with propensity score matching method,} \textit{BMJ open}, 10, e040062.

\bibitem[{Traskin and Small(2011)}]{traskin2011defining}
Traskin, M. and Small, D.~S. (2011), \enquote{Defining the study population for an observational study to ensure sufficient overlap: a tree approach,} \textit{Statistics in Biosciences}, 3, 94--118.

\bibitem[{Ward et~al.(2014)Ward, Alvarez, Vo, and Martin}]{ward2014direct}
Ward, A., Alvarez, P., Vo, L., and Martin, S. (2014), \enquote{Direct medical costs of complications of diabetes in the United States: estimates for event-year and annual state costs (USD 2012),} \textit{Journal of medical economics}, 17, 176--183.

\bibitem[{Yildirim and Wright(2002)}]{yildirim2002warm}
Yildirim, E.~A. and Wright, S.~J. (2002), \enquote{Warm-start strategies in interior-point methods for linear programming,} \textit{SIAM Journal on Optimization}, 12, 782--810.

\bibitem[{Yu and Rosenbaum(2019)}]{yu2019directional}
Yu, R. and Rosenbaum, P.~R. (2019), \enquote{Directional penalties for optimal matching in observational studies,} \textit{Biometrics}, 75, 1380--1390.

\bibitem[{Zhang and Zhao(2023)}]{zhang2023randomization}
Zhang, Y. and Zhao, Q. (2023), \enquote{What is a randomization test?} \textit{Journal of the American Statistical Association}, 1--29.

\bibitem[{Zubizarreta(2012)}]{zubizarreta2012using}
Zubizarreta, J.~R. (2012), \enquote{Using mixed integer programming for matching in an observational study of kidney failure after surgery,} \textit{Journal of the American Statistical Association}, 107, 1360--1371.

\bibitem[{Zubizarreta et~al.(2014)Zubizarreta, Paredes, and Rosenbaum}]{zubizarreta2014matching}
Zubizarreta, J.~R., Paredes, R.~D., and Rosenbaum, P.~R. (2014), \enquote{Matching for balance, pairing for heterogeneity in an observational study of the effectiveness of for-profit and not-for-profit high schools in Chile,} .

\bibitem[{Zubizarreta et~al.(2018)Zubizarreta, Zubizarreta, and SystemRequirements}]{zubizarreta2018package}
Zubizarreta, J.~R., Zubizarreta, M. J.~R., and SystemRequirements, G. (2018), \enquote{Package ‘designmatch’,} \textit{Matched Samples that are Balanced and Representative by Design Version 0.3. 0}.

\end{thebibliography}

\appendix

\section{\textcolor{black}{\texttt{two\_dist\_match}: tradeoffs between two pairwise distances}}
\label{subsec:twodist}
\textcolor{black}{The  analysis of Section \ref{subsec:distbal} focused on tradeoffs between one difficult-to-balance nominal variable and a multivariate distance incorporating all pre-treatment variables. 
Another common setting where tradeoff analysis is useful involves two distinct multivariate distances.  A good example is the tradeoff between matching on a propensity score, or an estimate of the probability of exposure as a function of pre-treatment covariates, and matching on a prognostic score, or an estimate of the outcome variable as a function of pre-treatment covariates (usually among controls alone).  Matching on propensity scores alone is common but involves some drawbacks \citep{lee2010improving,stuart2013prognostic}; in particular, propensity scores induce balance on covariates in expectation but may not produce finite-sample similarity between covariate distributions in practice. Furthermore, variables important in predicting treatment (the explicit focus of propensity score matching) may not be important in predicting outcomes, in which case propensity score matching may distract from more important goals in matching \citep{pimentel2016constructed, liao2023using}.  Matching on the prognostic score can remedy this particular issue \citep{hansen2008prognostic}.  While procedures have been suggested for matching on both propensity and prognostic scores \citep{antonelli2018doubly}, tradeoff analysis allows more fine-grained control of the similarity on both scores.}


\textcolor{black}{The \texttt{two\_dist\_match} command provided by \texttt{MultiObjMatch} can be used to investigate tradeoffs between any two multivariate distances.  We illustrate its use here to explore a tradeoff between propensity score matching and prognostic score matching in a simulated dataset.  300 independent observations are drawn, each with 12 mutually independent standard normal covariates $X_1, \ldots, X_{12}$.  Outcomes $Y$ and treatments $Z$ are generated according to the following models:
\begin{align*}
&Y = X_1  + X_2 + \epsilon \quad \quad \epsilon \sim \mathcal{N}(0,1)\\
&\text{logit}(P(Z=1|X_1, \ldots, X_{12}) = -2 + \frac{1}{2}\sum^6_{i=1}X_i.
\end{align*}
Note that there is no treatment effect since $Y$ is not a function of $Z$.  Code to generate this data is provided below.  The resulting dataset has 53 treated subjects and 247 controls.}
\begin{lstlisting}[language=R]
set.seed(2024 - 3 - 28)
n <- 300
p <- 12
X_w_pilot <- data.frame(matrix(rnorm(n * p), ncol = p))
colnames(X_w_pilot) <- paste("X", c(1:p), sep = "")
inv_logit <- function(x){
  1/(1 + exp(-x))
}
true_ps <- inv_logit(-2 + 0.5 * rowSums(X_w_pilot[, 1:6]))      
true_mu0 <- with(X_w_pilot, X1 + X2)
Z <- rbinom(n, size = 1, p = true_ps)
Y0 <- true_mu0 + rnorm(n)
data_w_pilot <- cbind(
                 as.data.frame(X_w_pilot), 
                 "Z" = Z, 
                 "Y0" = Y0
                )
\end{lstlisting}

\textcolor{black}{We next fit a propensity score and a prognostic score in the simulated data.  Due to potential biases arising from using in-sample outcomes in constructing matched samples, it is important that prognostic scores be fit in external samples \citep{rosenbaum1984consequences, hansen2008prognostic}; in contrast, the propensity score can be fit in-sample.  As such, in we take a random sample of 50 controls to dedicate for prognostic score estimation and use the full treatment group and the remaining 197  controls for the matched analysis.
The prognostic score is fit in the holdout sample under ordinary least square regression using all twelve covariates. The propensity score is fit in the main analysis sample using twelve covariates and a binary logistic regression model.}



\begin{lstlisting}[language=R]
n_pilot <- 50
pilot_rows <- sample(which(Z == 0), size = n_pilot)
pilot_data <- data_w_pilot[pilot_rows, ]
main_data <- data_w_pilot[-pilot_rows, ]

prog_score_lm <- lm(Y0 ~ . - Z, data = pilot_data)
main_data$prog_score <- predict(
                         prog_score_lm, 
                         newdata = main_data
                        )

prop_score_glm <- glm(
                    Z ~ . - Y0 - prog_score, 
                    data = main_data, 
                    family = "binomial"
                  )
main_data$prop_score <- fitted(prop_score_glm)
\end{lstlisting}
\textcolor{black}{ A distance matrix is then constructed for each score, with an element $(i,j)^{th}$ equal to the absolute difference between $i^{th}$ treatment unit and $j^{th}$ control unit's scores.}
\begin{lstlisting}[language=R]
distance_prognostic <- as.matrix(dist(main_data$prog_score))
idx <- 1:nrow(main_data)
treatedUnits <- idx[main_data$Z == 1]
controlUnits <- idx[main_data$Z == 0]
d_prognostic <- distance_prognostic[treatedUnits, 
                                    controlUnits]

distance_pscore <- as.matrix(dist(main_data$prop_score))
d_propensity <- distance_pscore[treatedUnits, controlUnits]
\end{lstlisting}
\textcolor{black}{Next we consider an optimal propensity score match and consider the quality of balance obtained.}
\begin{lstlisting}[language = R]
ps_match <- matchit(
             Z ~ . - Y0 - prog_score,
             data = main_data,
             method = "optimal", 
             distance = "glm"
            )
ps_bal <- bal.tab(
           ps_match,
           data = main_data, 
           quick = FALSE,
           s.d.denom = "pooled"
          )
subset(ps_bal$Balance, select = c("Diff.Un", "Diff.Adj"))
\end{lstlisting}
\textcolor{black}{Although the propensity score match balances most variables reasonably well it leaves absolute standardized differences larger than 0.15 on five covariates (Table \ref{tab:prog_balance_improve}). It is reasonable to try to improve balance further, but a balance constraint on a single variable is not an obvious solution as it was in the job training example of Section \ref{subsec:distbal} or the DKD example of Section \ref{sec:dkd}, since multiple variables exhibit similar moderate imbalance.  Instead, we will introduce similarity on the prognostic score within pairs as a second design objective, with the goal of addressing residual imbalances specifically for those variables with strong outcome correlations.}

\begin{figure}
\begin{tabular}{cc}
-\vspace{1em}
(a) \includegraphics[scale=0.3]{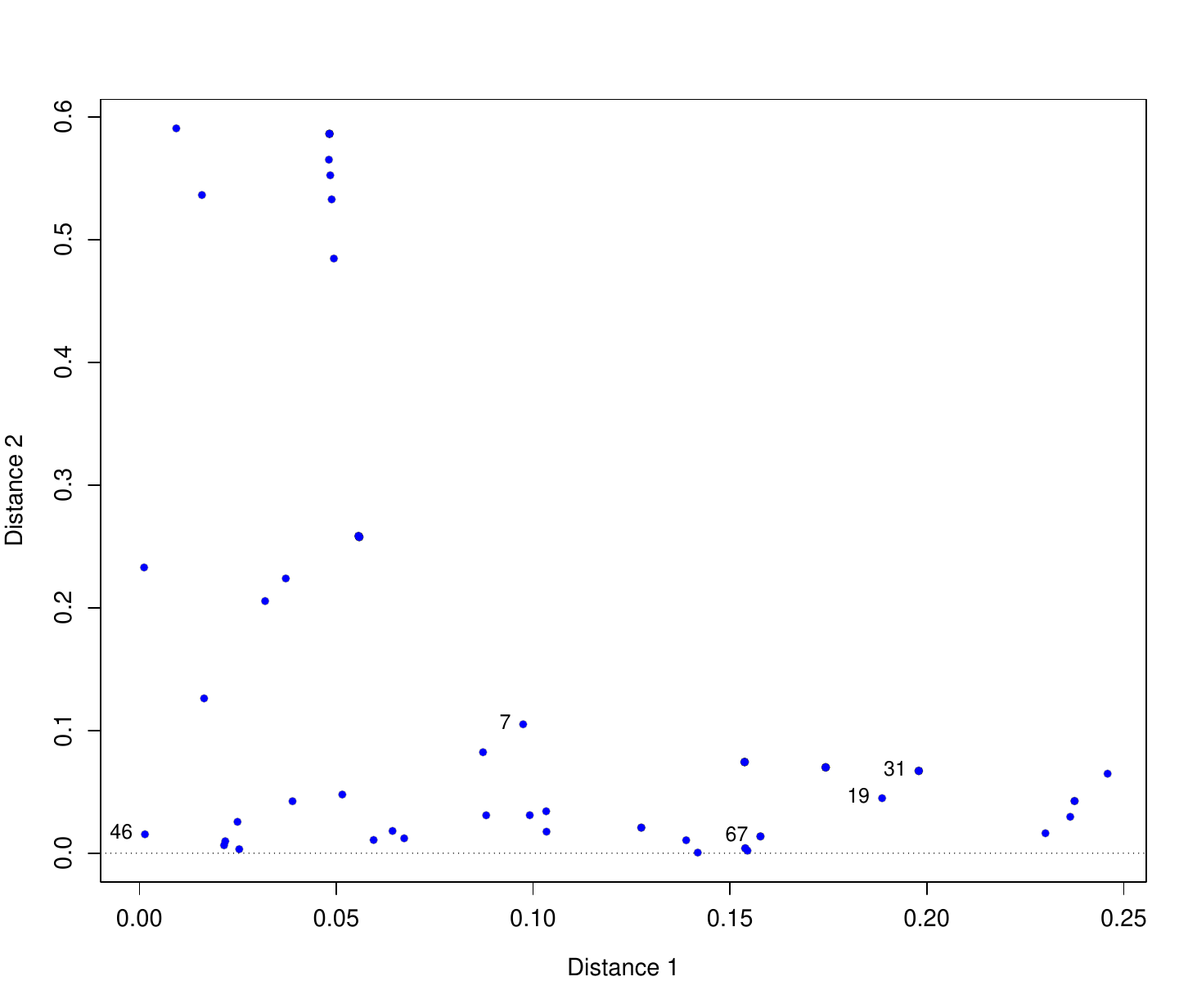} & (b) \includegraphics[scale=0.45]{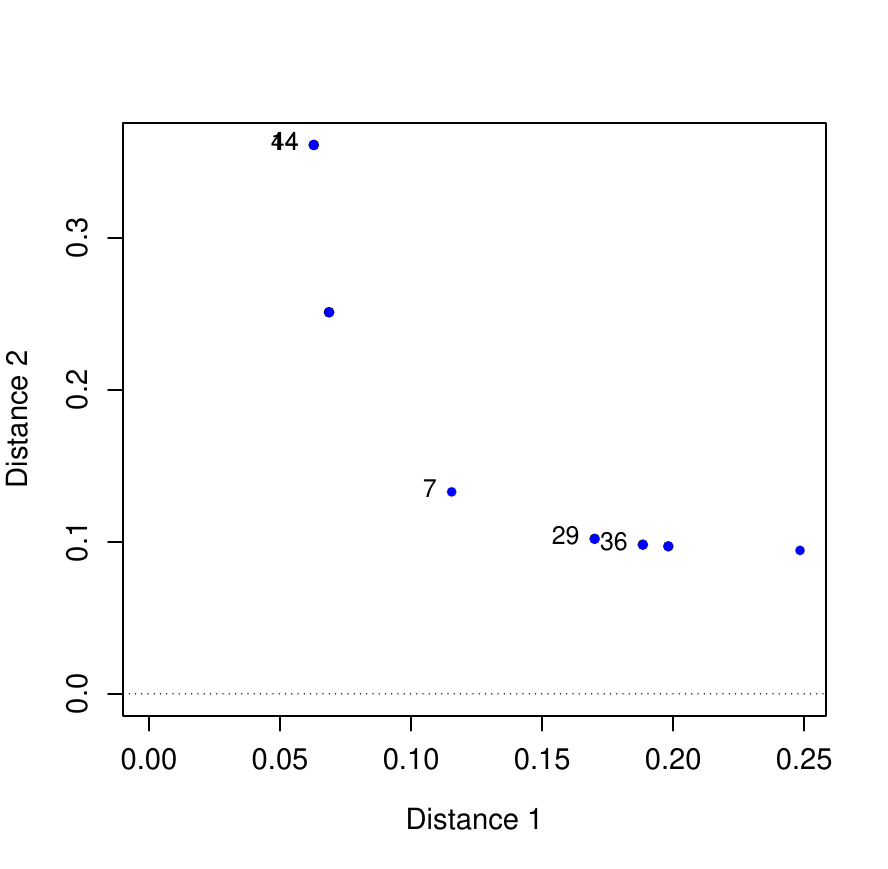} 
\end{tabular}
\caption{  \textcolor{black}{Tradeoff plots for prognostic score vs. propensity score matching (a) using all solutions from the initial grid search and (b) selecting only those matches with no treated subjects excluded.}}
\label{fig:two_dist_figs}
\end{figure}

\textcolor{black}{We conduct a range of matches using the package command \texttt{two\_dist\_match}, varying the relative emphasis on the propensity and prognostic scores.  
Here we specify the \texttt{dist1\_type} and \texttt{dist2\_type} arguments as `user' since we built the propensity and prognostic score distances already.  However, \texttt{two\_dist\_match} also supports the option to compute certain distances internally by specifying \texttt{dist1\_type = "euclidean"} or \texttt{dist1\_type = "robust\_mahalanobis"} and providing in argument \texttt{dist1\_col} a vector of covariate names (in the data frame that must in this case be passed to the \texttt{data} argument) to use for the distance computation.}  
\begin{lstlisting}[language=R]
prop_prog <- two_dist_match(
                                dist1_type = "user", 
                                dist2_type = "user",
                                dist1_matrix = d_propensity, 
                                dist2_matrix = d_prognostic,
                                max_iter = 1
                               ) 
\end{lstlisting}
\textcolor{black}{As in Section \ref{subsec:distbal}, numerical diagnostics and visualizations can be produced easily by functions \texttt{summary} and \texttt{visualize}.  Note that we use the ``average" argument to the \texttt{visualize} command to present averages across matched pairs (rather than sums) on both axes for clearer interpretability.  Table \ref{tab:prog_prop_zero_exclude} shows numbers produced by the \texttt{summary} command for the individual matched solutions in Figure \ref{fig:two_dist_figs}(b), and Table \ref{tab:prog_balance_improve} shows the SMDs for the ten initial covariates for a group of selected matches in comparison to the original propensity score match.  A tradeoff plot showing all solutions computed by the initial grid search  is shown in Figure \ref{fig:two_dist_figs}(a).  Since we only have 53 treated units to begin with and the initial propensity score match is already reasonably good, we restrict attention to matches that exclude no treated subjects (Figure \ref{fig:two_dist_figs}(b)). }
\begin{lstlisting}[language=R]
visualize(prop_prog, average_cost = TRUE)
no_exclusion <- filter_match_result(prop_prog, 
                                    "fExclude == 0")
visualize(no_exclusion, average_cost = TRUE)
\end{lstlisting}



\begin{table}[h!]
  \textcolor{black}{
\begin{tabular}{l d{1} d{1.3} d{1} d{1.3} d{1} d{1.3}}
\hline
Matching & \multicolumn{1}{c}{Prop. Score} & \multicolumn{1}{c}{Exclusion} & \multicolumn{1}{c}{Prog. Score} & \multicolumn{1}{c}{Prop. Score} & \multicolumn{1}{c}{Exclusion} & \multicolumn{1}{c}{Prog. Score}  \\ 
Index & \multicolumn{1}{c}{Penalty} & \multicolumn{1}{c}{Penalty} & \multicolumn{1}{c}{Penalty} & \multicolumn{1}{c}{Distance} & \multicolumn{1}{c}{Cost} & \multicolumn{1}{c}{Distance} \\ \hline 
  44 & 1 & 95.35 & 0.01 & 3.21 & 0 & 18.41 \\ 
  10 & 1 & 953.51 & 0.10 & 3.50 & 0 & 12.80 \\ 
  7 & 1 & 9.54 & 1.00 & 5.89 & 0 & 6.78 \\ 
  29 & 1 & 32.42 & 3.40 & 8.67 & 0 & 5.20 \\ 
  36 & 1 & 63.89 & 6.70 & 9.61 & 0 & 5.00 \\ 
  13 & 1 & 953.51 & 10.00 & 10.10 & 0 & 4.95 \\ 
  9 & 1 & 953.51 & 100.00 & 12.67 & 0 & 4.81 \\ \hline
\end{tabular}}


\caption{\textcolor{black}{Tabular summary of tradeoffs between prognostic score and propensity score matching in the simulaated data example without excluding subjects, via  function \texttt{summary}. Additional solutions with identical values for both objective functions (to two decimal places) are excluded from the table.}}
\label{tab:prog_prop_zero_exclude}
\end{table}
\begin{table}[ht]
\centering
\small
  \textcolor{black}{
\begin{tabular}{r|r|rrrrrrr}
  \hline
  Cov- & PS & \multicolumn{7}{c}{Index of match in tradeoff solution set}\\
 ariate  & match & 44 & 10 & 7 & 29 & 36 & 13 & 9 \\ 
  \hline
X1 & 0.08 & 0.05 & 0.01 & 0.01 & 0.04 & -0.01 & 0.02 & 0.11 \\ 
  X2 & -0.09 & -0.03 & -0.05 & 0.03 & 0.00 & 0.03 & 0.00 & -0.10 \\ 
  X3 & -0.09 & -0.06 & -0.15 & -0.23 & -0.11 & -0.07 & -0.09 & -0.15 \\ 
  X4 & 0.20 & 0.14 & 0.20 & 0.35 & 0.46 & 0.50 & 0.56 & 0.66 \\ 
  X5 & 0.10 & 0.04 & 0.15 & -0.02 & 0.19 & 0.22 & 0.24 & 0.24 \\ 
  X6 & 0.04 & 0.13 & 0.08 & 0.26 & 0.32 & 0.36 & 0.45 & 0.65 \\ 
  X7 & -0.01 & -0.02 & -0.05 & 0.05 & 0.21 & 0.15 & 0.12 & 0.19 \\ 
  X8 & -0.08 & -0.04 & -0.03 & 0.02 & -0.14 & -0.16 & -0.05 & 0.04 \\ 
  X9 & 0.00 & -0.09 & -0.09 & -0.02 & 0.02 & 0.08 & 0.07 & 0.09 \\ 
  X10 & 0.01 & 0.07 & 0.01 & 0.11 & 0.17 & 0.15 & 0.12 & 0.24 \\ 
  X11 & -0.08 & -0.02 & -0.02 & 0.10 & 0.05 & -0.01 & 0.00 & -0.09 \\ 
  X12 & 0.11 & 0.08 & 0.04 & -0.06 & -0.06 & -0.15 & -0.17 & -0.28 \\ 
  prop\_score & 0.29 & 0.30 & 0.31 & 0.56 & 0.78 & 0.85 & 0.94 & 1.15 \\ 
   \hline
\end{tabular}}
\caption{  \textcolor{black}{SMDs on all twelve covariates and the propensity score for no-exclusion matches in the simulated example, including both the initial optimal propensity score match and a range of matches computed by the tradeoff algorithm.}}
\label{tab:prog_balance_improve}
\end{table}

  \textcolor{black}{From Tables \ref{tab:prog_prop_zero_exclude}-\ref{tab:prog_balance_improve}, the matches with indices 44 and 10 jump out as being particularly attractive.  Both have similar standardized differences to the propensity score match but do a little better overall, particularly on the prognostically important covariates X1 and X2.  
  In short, without excluding units or detracting substantially from quality of balance on most variables, tradeoff analysis has helped us improve on the already-moderately-good match available through optimal propensity score matching alone.}


\section{  \textcolor{black}{Choosing penalties for the DKD example} }\label{sec:dkd_hyperparam_searching}


  \textcolor{black}{We now elaborate on the process of exploring penalty combinations for the DKD example in Section \ref{sec:dkd}. The DKD dataset cannot be shared publicly in light of patient confidentiality requirements, but we provide key R commands to illustrate our analysis. 
We begin with an automatic grid search. Here \texttt{treat\_val} contains the name of the treatment variable in the original dataframe \texttt{dkd}, \texttt{bal\_val} contains the name of the microvascular complications variable, and \texttt{covariates} contains the names of all covariates deemed relevant. As discussed in Section \ref{sec:dkd}, a propensity score caliper is also imposed.  The propensity score is fit using logistic regression model on the covariates specified in \texttt{propensity\_col}, with caliper size (in standardized deviations of estimated propensity scores)  using \texttt{caliper\_option}.   The resulting set of matches includes a large number with 700 or more DKD patients excluded (over 20\% of the treated sample).  Since it is possible to achieve acceptable values of the other objective functions without excluding this many subjects (as shown below), we begin by removing all matches with 700 or more exclusions from consideration.  The \texttt{filter\_match\_result} command provides a convenient way to do this.}
\begin{lstlisting}[language=R]
dkd_initial_match <- dist_bal_match(
  dkd, 
  treat_col = treat_val,
  marg_bal_col = bal_val, 
  dist_col  = covariates,
  exact_col = "year", 
  propensity_col = covariates,
  caliper_option = 0.25
)

filtered_dkd_matching <- filter_match_result(
  dkd_initial_match,
  filter_expr = "fExclude < 700"
)
\end{lstlisting}

\begin{figure}[H]
    \centering
    \includegraphics[width=\textwidth]{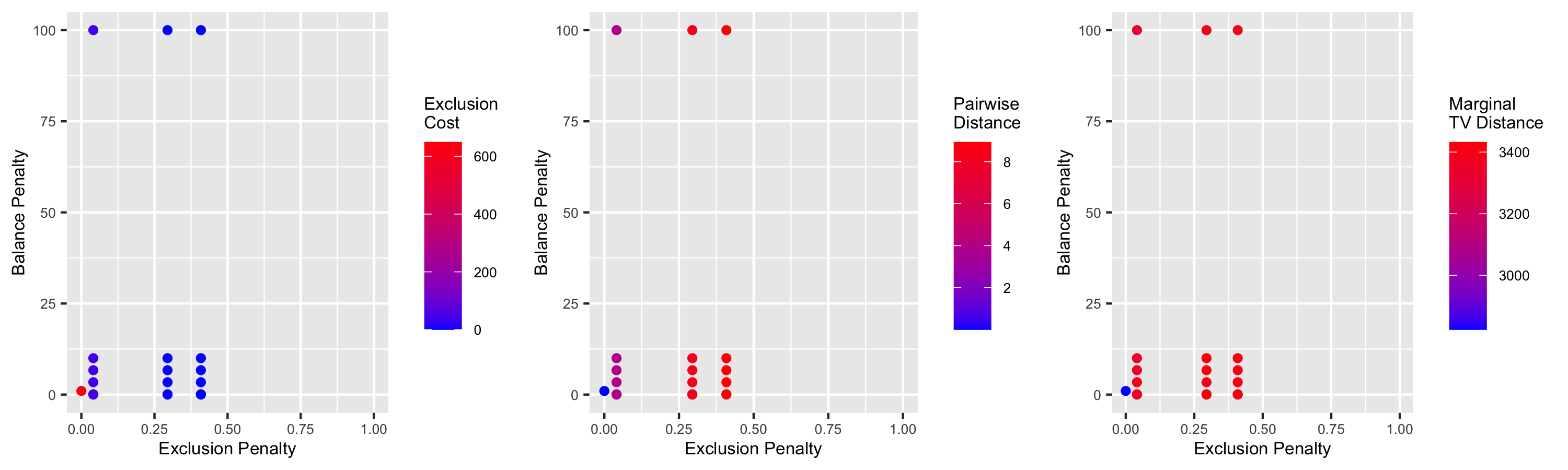}
    \caption{  \textcolor{black}{Objective function values over the 2-dimensional space of penalties for matches with less than 700 units dropped in the DKD example.}}
    \label{fig:dkd_param_search}
\end{figure}

  \textcolor{black}{As shown in Figure \ref{fig:dkd_param_search}, meaningful tradeoffs appear when the marginal balance penalty is 1 and the exclusion penalty increases from 0 to small values below 1. Therefore, we fix the balance penalty at 1 and vary the exclusion penalties over a range of values between 0 and 0.5. As illustrated above in Section \ref{sec:dkd}, the resulting matches represent a wide range of interesting solutions, dropping as few as 0 and as many as 581 DKD patients.  }

\begin{lstlisting}[language=R]
rho_balance <- 1 
rho_exclusion <- c(0.0005, 0.001,0.003, 0.005,0.008,
                   0.01, 0.02, 0.03,0.1,0.5)
\end{lstlisting}

\section{R console output from \texttt{summary} command}
\label{sec:console_output}

\begin{lstlisting}[language=R]
> summary(match_result)
[1] "rhoPair: penalty for distance objective function;rhoExclude: penalty for exclusion cost;rhoMarginal: penalty for marginal balance."
[1] "fPair: pair-wise distance sum objective function;fExclude: number of treated units left unmatched;fMarginal: marginal imbalance measured as the total\n        variation distance on the marginal distribution of speicified variable."
   match_index rhoPair rhoExclude rhoMarginal    fPair fExclude fMarginal
1            1       1      94.27        85.0 548.8697        0       138
2            2       1      94.27        86.0 540.6121        1       136
3            3       1      94.27        86.5 524.3096        3       132
4            4       1      94.27        87.0 501.7489        6       126
5            5       1      94.27        87.5 480.6461        9       120
6            6       1      94.27        88.0 428.8319       17       104
7            7       1      94.27        88.5 351.3027       30        78
8            8       1      94.27        89.0 284.8186       42        54
9            9       1      94.27        90.0 193.2918       61        16
10          10       1      94.27        91.0 160.5134       69         0


> summary(match_result, type = "balance")
               type           1           3            5            7           10
age         Contin. -0.10407280 -0.13036732 -0.134415520 -0.108404098 -0.159532872
educ        Contin. -0.01375679    0.01419353    0.014520692 -0.002693724  0.028041945
married      Binary -0.12432432 -0.12087912 -0.113636364 -0.070967742 -0.034482759
nodegree     Binary   0.04864865   0.03846154    0.028409091  0.032258065    0.043103448
race_black   Binary   0.37297297   0.36263736    0.340909091  0.251612903    0.000000000
race_hispan  Binary -0.02162162 -0.01648352 -0.005681818 -0.006451613  0.000000000
race_white   Binary -0.35135135 -0.34615385 -0.335227273 -0.245161290  0.000000000
re74        Contin. -0.13870789 -0.13190865 -0.125649477 -0.098354004 -0.003280656
re75        Contin. -0.04873460 -0.03856235 -0.035913472 -0.007374445  0.042001758


> summary(match_result, type = "exclusion")
   Matching Index Number of Matched Units Percentage of Matched Units
1               1                     185                   1.0000000
2               2                     184                   0.9945946
3               3                     182                   0.9837838
4               4                     179                   0.9675676
5               5                     176                   0.9513514
6               6                     168                   0.9081081
7               7                     155                   0.8378378
8               8                     143                   0.7729730
9               9                     124                   0.6702703
10             10                     116                   0.6270270
\end{lstlisting}

\end{document}